\documentclass[varenna]{cimento}

\def\kpnnp{$K^+ \to \pi^+ \nu\bar\nu$}
\def\bkpnnp{$B(K^+ \to \pi^+ \nu\bar\nu)$}
\def\kpnn0{$K_L\to\pi^0\nu\bar\nu$}
\def\bkpnn0{$B(K_L\to\pi^0\nu\bar\nu)$}
\def\kmm{$K_L\to\mu^+\mu^-$}
\def\bkmm{$B(K_L\to\mu^+\mu^-)$}

\def\kp0{$K_L \to \pi^0 \pi^0$}

\def\bkppr{B(K_L \to \pi^+ \pi^-)}
\def\kpll{$K_L \to \pi^0 \ell^+ \ell^-$}
\def\kpmm{$K_L \to \pi^0 \mu^+ \mu^-$}
\def\klpee{$K_L \to \pi^0 e^+ e^-$}
\def\kpgg{$K_L \to \pi^0 \gamma\gamma$}

\def\be{\begin{equation}}
\def\ee{\end{equation}}
\def\bea{\begin{eqnarray}}
\def\eea{\end{eqnarray}}

\usepackage{graphicx}  
\title{Rare Kaon Decays}
\author{L.~Littenberg
 }
\institute{Brookhaven National Laboratory, Upton, N.Y. 11973}
\PACSes{\PACSit{43.35.Ei}\PACit{78.60.Mq}}

\begin{document}

\maketitle

%

\section{Introduction}

In recent years the study of rare kaon decays has had three primary 
motivations: 
\begin{enumerate}
\item The search for physics beyond the Standard Model (BSM).  Virtually
all attempts to improve on the Standard
Model (SM) predict some degree of lepton flavor violation (LFV).
Decays such as $K_L \to \mu^{\pm} e^{\mp}$ and $K^+ \to \pi^+ \mu^+
e^-$ have excellent experimental signatures and can consequently be
pursued to sensitivities that correspond to extremely high 
energy scales in models where there can be a tree-level contribution 
and the only suppression is that of the mass of the exchanged 
field.  There are also theories that predict new particles created in 
kaon decay or the violation of symmetries other than lepton flavor.
\\
\item The potential of decays that are allowed 
but that are very suppressed in the SM.  In a number of kaon decays,
the leading component
is a G.I.M.-suppressed\cite{Reference:GIM} one-loop process that 
is quite sensitive to fundamental SM parameters such as $V_{td}$.
Because of the severe suppression,
these decays are also potentially very sensitive to BSM physics.
\\
\item Long-distance-dominated decays can
test theoretical techniques such as chiral perturbation theory
($\chi$PT) that attempt to account for the low-energy behavior of QCD.
Also, information from some of these decays is required to extract
fundamental information from certain of the one-loop processes.
\end{enumerate}

        This field remains active, as indicated by Table \ref{decays:a},
that lists the rare 
decays for which results have emerged in the last few years,
as well as those that are under analysis. It is clear that
one must be quite selective in a review of moderate length.

\begin{table}[h]
\caption{\it Rare $K$ decay modes under recent or on-going study.}
\begin{tabular}{llll} 
\hline
$K^+ \to \pi^+ \nu\bar\nu$ & $K_L \to \pi^0 \nu\bar\nu$ &
$K_L \to \pi^0 \mu^+\mu^-$ & $K_L \to \pi^0 e^+e^-$ \\
$K^+ \to \pi^+ \mu^+\mu^-$ & $K^+ \to \pi^+ e^+e^-$ &
$K_L \to  \mu^+\mu^-$ & $K_L \to  e^+e^-$ \\
$K^+ \to \pi^+ e^+ e^- \gamma$ & $K^+ \to \pi^+ \pi^0 \nu\bar\nu$ &
$K_L \to e^{\pm} e^{\mp} \mu^{\pm} \mu^{\mp}$ & $K^+ \to \pi^+ \pi^0 \gamma$ \\
$K_L \to \pi^+ \pi^- \gamma$ & $K_L \to \pi^+ \pi^- e^+ e^-$ &
$K^+ \to \pi^+ \pi^0 e^+ e^-$ & $K^+ \to \pi^0 \mu^+ \nu \gamma$ \\
$K_L \to \pi^0 \gamma \gamma$ & $K^+ \to \pi^+ \gamma \gamma$ &
$K^+ \to \mu^+ \nu \gamma$ & $K^+ \to e^+ \nu e^+ e^-$ \\
$K^+ \to \mu^+ \nu e^+ e^-$ & $K^+ \to e^+ \nu \mu^+\mu^-$ &
$K_L \to e^+ e^- \gamma$ & $K_L \to \mu^+ \mu^- \gamma$ \\
$K_L \to e^+ e^- \gamma\gamma$ & $K_L \to \mu^+ \mu^- \gamma\gamma$ &
$K_L \to e^+ e^- e^+ e^-$ & $K_L \to \pi^0 e^+ e^- \gamma$ \\
$K_S \to \pi^0 e^+ e^-$ & $K_S \to \pi^0 \mu^+ \mu^-$ & 
$K_S \to \gamma\gamma$ & $K_S \to \pi^0 \gamma \gamma$ \\
$K^+ \to \pi^+ \mu^+e^-$ & $K_L \to \pi^0 \mu^{\pm} e^{\mp}$ &
$K_L \to \mu^{\pm} e^{\mp}$ & $K^+ \to \pi^- \mu^+ e^+$ \\
$K^+ \to \pi^- e^+ e^+$ & $K^+ \to \pi^- \mu^+ \mu^+$ &
$K^+ \to \pi^+ X^0$ & $K_L \to e^{\pm} e^{\pm} \mu^{\mp} \mu^{\mp}$ \\
$K^+ \to \pi^+ \gamma$& $K_L \to \pi^0 \pi^0 e^+ e^-$ &
$K_L \to \pi^0 \pi^0 \nu\bar\nu$ & $K_L \to \pi^+ \pi^- \pi^0 e^+ e^-$ \\
\hline
\end{tabular}
\label{decays:a}
\end{table}

\section{Beyond the Standard Model}

\begin{figure}[h]
\centering
 \includegraphics[angle=0, height=.24\linewidth]{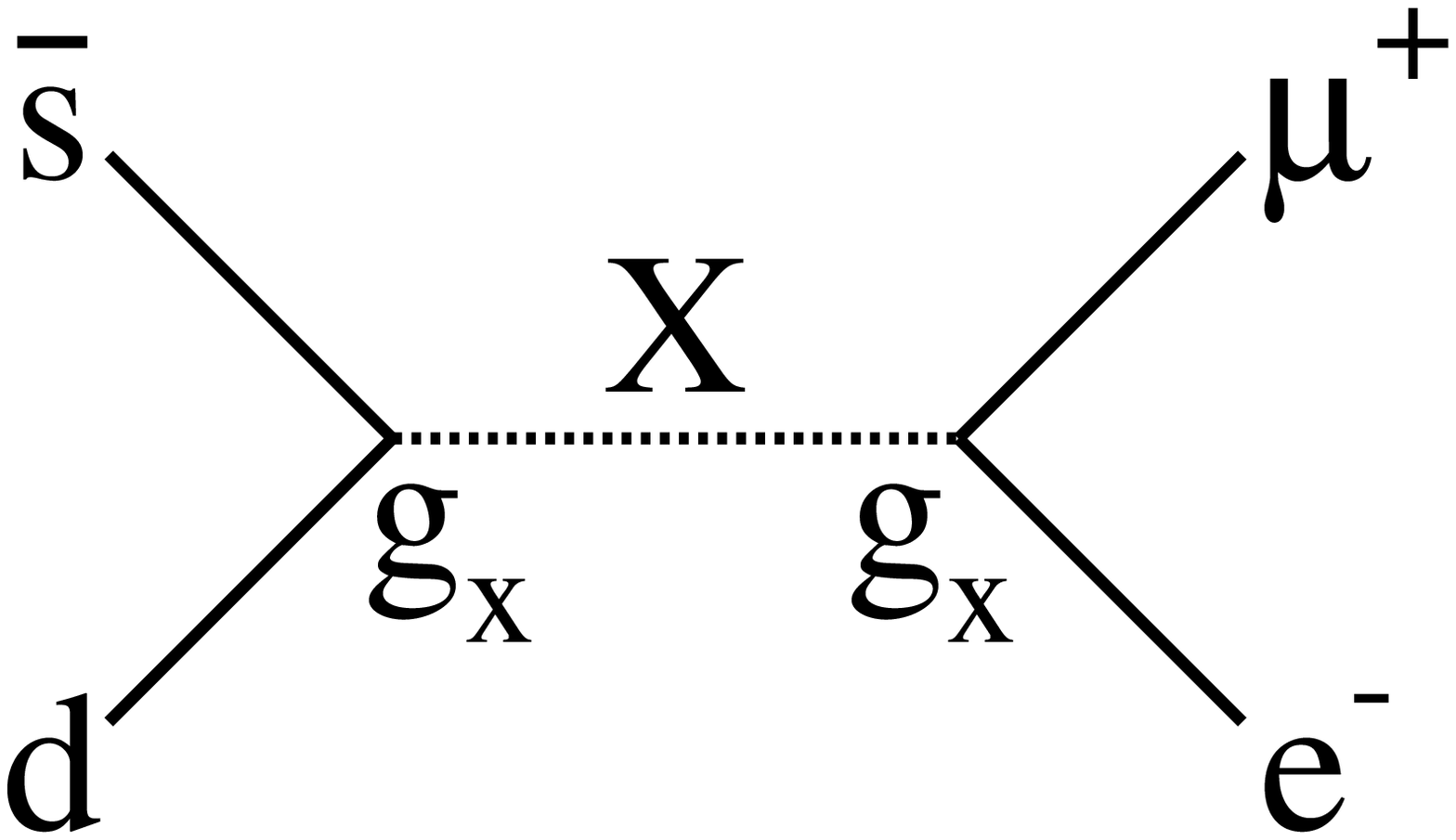}
 \includegraphics[angle=0, height=.24\linewidth]{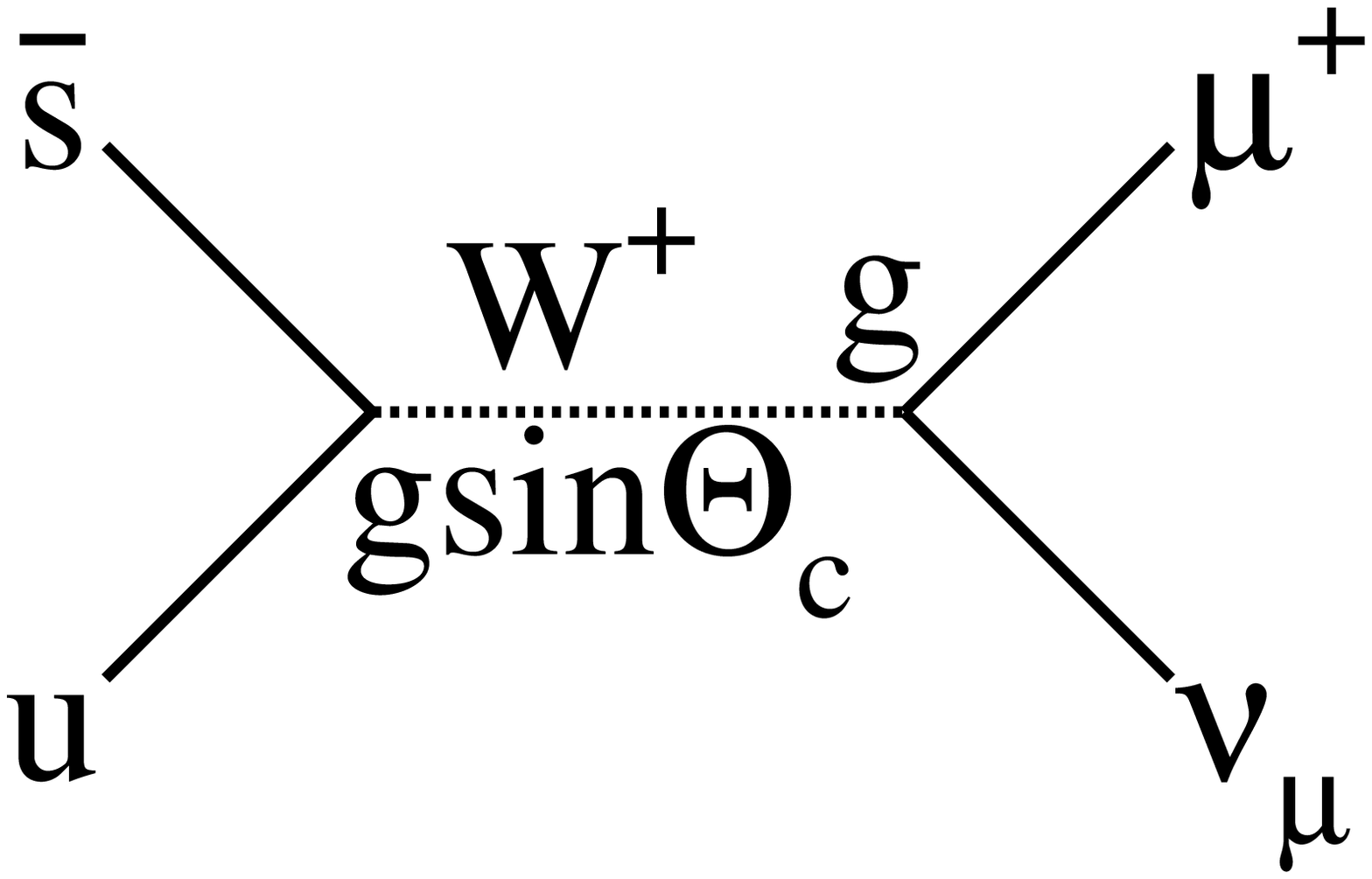}
  \caption{Horizontal gauge boson mediating $K_L \to \mu e$, compared with
$W$ mediating $K^+ \to \mu^+ \nu$.
    \label{fig:poster} }
\end{figure}

	The poster children for BSM probes in kaon decay are LFV
processes like $K_L \to \mu e$ and $K^+ \to \pi^+ \mu^+ e^-$.  In
principle, these can proceed through neutrino mixing, but the known
neutrino mixing parameters limit the rate through this mechanism to a
completely negligible level~\cite{Lee:1977ti}.  Thus the observation of
LFV in kaon decay would require a new mechanism.
Fig.\ref{fig:poster} shows $K_L \to \mu e$ mediated by a
hypothetical horizontal gauge boson $X$, compared with the
kinematically very similar process $K^+ \to \mu^+ \nu$ mediated by a
$W$ boson.

\noindent
Using measured values for $M_W$, the $K_L$ and $K^+$ decay rates and
$B(K^+ \to \mu^+ \nu)$, and assuming a $V-A$ form for the
new interaction, one can show~\cite{Cahn:1980kv}:
\be
M_X \approx 220\, TeV/c^2 \left[{g_X \over{g}}\right]^{1/4} \left[ {10^{-12} \over{B(K_L \to \mu e)}} \right]^{1/4}
\label{LFVcomp}
\ee so that truly formidable scales can be probed if $g_X \sim g$
(see also \cite{Rizzo:1998fv}).  In addition
to this generic picture, there are specific models, such as extended
technicolor in which LFV at observable levels in kaon decays is 
quite natural~\cite{Eichten:1986eq}.

        There were a number of $K$ decay experiments primarily
dedicated to lepton flavor violation at the Brookhaven AGS during the
1990's.  These advanced the sensitivity to such processes by many
orders of magnitude.  In addition, several ``by-product'' results on
LFV and other BSM topics have emerged from the other kaon decay
experiments of this period.  Rare kaon decay experiments often also
yield results on $\pi^0$ decays, since these can readily be tagged,
{\it e.g.} via $K^+ \to \pi^+ \pi^0$ or $K_L \to \pi^+ \pi^- \pi^0$.
Table~\ref{BSM} summarizes the status of work on BSM probes in kaon
decay.  The relative reach of these processes is best assessed
by comparing the partial rates rather than the branching ratios.

\begin{table}[h]
\caption{\it Current 90\% CL limits on  $K$ decay modes violating the SM.  The violation
codes are ``LF'' for lepton flavor, ``LN'' for lepton number,
``G'' for generation number, \protect\cite{Cahn:1980kv}, ``H'' for helicity, ``N'' requires new particle}
\begin{tabular}{|c|c|c|c|c|c|} \hline
Process & Violates & 90\% CL BR Limit & $\Gamma$ Limit (sec$^-1$) & Experiment & Reference \\
\hline
\hline
$K_L \to \mu e $ & LF& $4.7 \times 10^{-12}$ & $9.1 \times 10^{-5}$ &AGS-871 & \cite{Ambrose:1998us} \\
$K^+ \to \pi^+ \mu^+ e^-$ & LF &  $1.2 \times 10^{-11}$ & $9.7 \times 10^{-4}$ & AGS-865 &\cite{Sher:2005sp}  \\
$K^+ \to \pi^+ \mu^- e^+$ & LF, G &$5.2 \times 10^{-10}$ & $4.2 \times 10^{-2}$ & AGS-865 & \cite{Appel:2000tc} \\
$K_L \to \pi^0 \mu e$ & LF & $3.37 \times 10^{-10}$ & $6.4 \times 10^{-3}$ &  KTeV & \cite{Bellavance:2003uu}\\
$K^+ \to \pi^- e^+ e^+$ & LN, G &$6.4 \times 10^{-10}$ & $5.2 \times 10^{-2}$ & AGS-865 & \cite{Appel:2000tc} \\
$K^+ \to \pi^- \mu^+ \mu^+$ & LN, G &$3.0 \times 10^{-9}$ & $2.4 \times 10^{-1}$ & AGS-865 & \cite{Appel:2000tc} \\
$K^+ \to \pi^- \mu^+ e^+$ & LF, LN, G &$5.0 \times 10^{-10}$ & $4.0 \times 10^{-2}$ & AGS-865 & \cite{Appel:2000tc} \\
$K_L \to \mu^{\pm} \mu^{\pm} e^{\mp} e^{\mp} $ & LF, LN, G & $4.12 \times 10^{-11}$ & $8.0 \times 10^{-4}$ & KTeV & \cite{Alavi-Harati:2002eh}\\
$K^+ \to \pi^+ f^0$ & N & $5.9 \times 10^{-11}$ & $4.8 \times 10^{-3}$ & AGS-787 & \cite{Adler:2001xv}\\
$K^+ \to \pi^+ \gamma$ & H & $2.3 \times 10^{-9}$& $1.9 \times 10^{-1}$ & AGS-949 & \cite{Artamonov:2005ru}\\
\hline
\end{tabular}
\label{BSM}
\end{table}

This table makes it clear that any deviation from the SM must be
highly suppressed.  In a real sense the kaon LFV probes have become the
victims of their own success.  By and large the particular theories they were
designed to probe have been forced to retreat to
the point where meaningful tests in the kaon system would be very
difficult (although there are exceptions\cite{Appelquist:2004ai}). 
The currently more popular theoretical approaches tend to predict
a rather small degree of LFV in kaon decays.
For example, although these decays do provide the most
stringent limits on strangeness-changing R-violating couplings, the
minimal supersymmetric extension of the Standard Model (MSSM) predicts LFV in
kaon decay at levels far beyond the current experimental state of the
art \cite{Belyaev:2000xt}.  Decays such as $K^+ \to \pi^- \mu^+ \mu^+$,
that violate lepton number as well as flavor, are possible in the MSSM,
but even more drastically suppressed\footnote{There are models 
involving sterile neutrinos in which such processes are conceivably 
observable\cite{Dib:2000wm}.}.

\begin{figure}[t]
\centering
 \includegraphics[angle=0, height=.65\textheight]{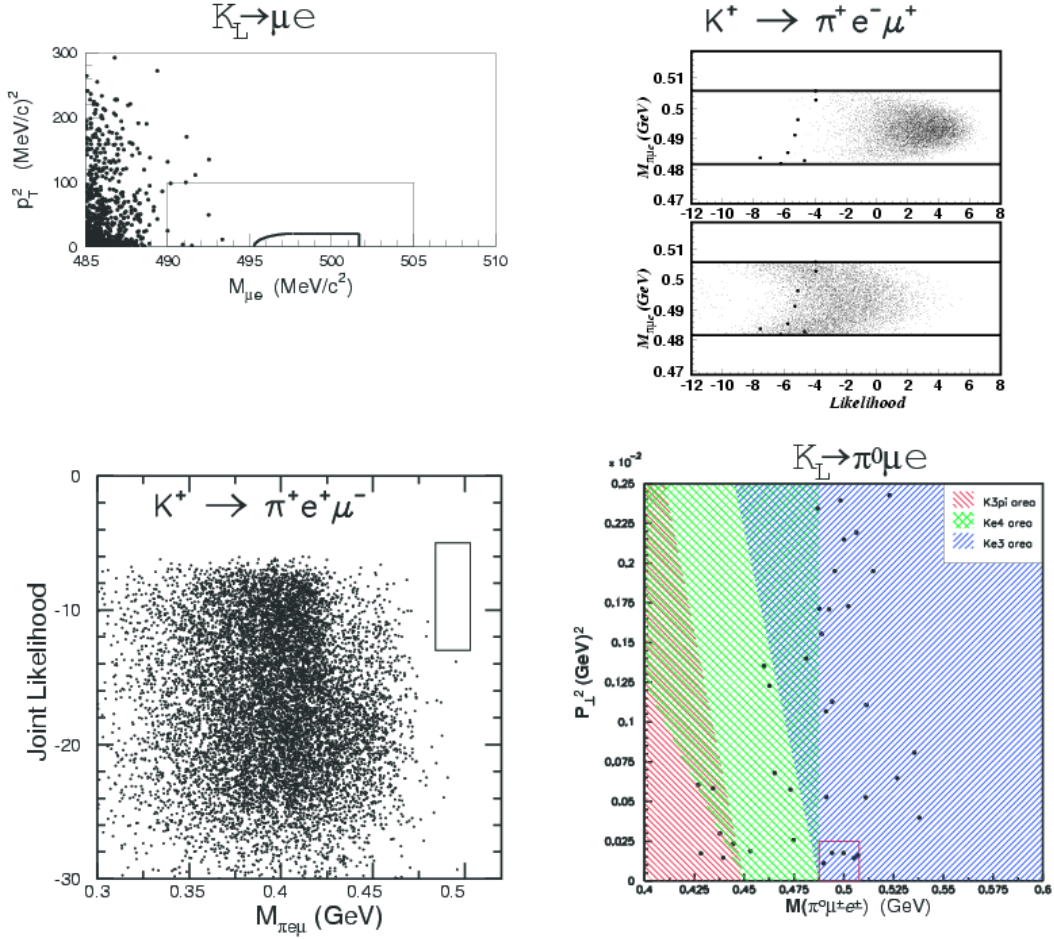}
  \caption{Signal planes showing candidates for LFV kaon decays from
recent experiments.
{\bf Top left:} $p_T^2$ vs $M_{\mu e}$ from Ref.~\protect\cite{Ambrose:1998us}, 
{\bf Top right:} $M_{\pi \mu e}$ vs Log likelihood from Ref.~\protect\cite{Sher:2005sp}
(upper plot shows signal Monte Carlo, lower plot shows background Monte Carlo,
observed events are shown as larger dots on both), 
{\bf Bottom left:} Joint likelihood vs $M_{\pi \mu e}$ from Ref.~\protect\cite{Appel:2000tc},
and {\bf Bottom right:} $P^2_{\perp}$ vs $M_{\pi \mu e}$ from Ref.~\protect\cite{Bellavance:2003uu}.
    \label{fig:boxes} }
\end{figure}

	There have been some recent exceptions to the waning of
theoretical interest in kaon LFV, including models with extra dimensions
\cite{Frere:2004yu}, but even in an improved motivational climate, 
there would be barriers to rapid future progress.  Although K fluxes
significantly greater than those used in the last round of LFV experiments 
are currently available, the rejection of background is a significant
challenge.   Fig.\ref{fig:boxes} shows the signal planes
of four of the most sensitive LFV searches.  It is clear that
background either is already a problem or soon would be if the statistical
sensitivity of these searches were increased.  Thus new techniques
will need to be developed to push such searches significantly
further.

At the moment no new kaon experiments focussed on LFV are being
planned.  Interest in probing LFV has largely migrated to the muon
sector.

One exception to the poor prospects for dedicated BSM searches in kaon
decay is the search for $T$-violating (out-of-plane) $\mu^+$ polarization in $K^+ \to
\pi^0 \mu^+ \nu$ \footnote{$K\mu 3$ is not rare (BR $\sim 3\%$), 
but the amplitude being probed in this type of experiment is quite
small.  If the decay went entirely through this amplitude, at the
current limit, $|{\rm Im}(\xi ) | < 0.016$ at 90\% C.L.\cite{Imazato:2004aq} 
($\xi\equiv f_-/f_+$ where $f_{\pm}$ are the form factors of $K\mu 3$),
this would correspond to a branching ratio of $\sim 10^{-7}$. }.  There's a Letter 
of Intent\cite{jparc19} to continue the work of the current experiment, KEK 
E246, at the J-PARC facility currently under construction.  The proponents
seek to make an order of magnitude advance in precision on the measurement
of the polarization.  Since this is an interference effect the advance is 
roughly equivalent to that of two orders of magnitude in BR sensitivity.
This measurement is quite sensitive to BSM physics, particularly multi-Higgs 
models including certain varieties of supersymmetry\cite{Garisto:1991fc,Belanger:1991vx,
Kobayashi:1995cy,Fabbrichesi:1996eb,Wu:1996zq}.

\section{One-loop decays}

\begin{figure}[h]
\centering
 \includegraphics[angle=0, height=.08\textheight]{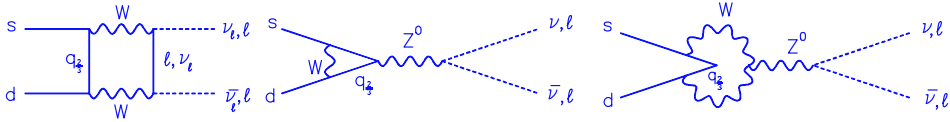}
  \caption{
Loop contributions to K decays
    \label{fig:loops} }
\end{figure}

\begin{figure}[h]
\centering
 \includegraphics[angle=90, height=.3\textheight]{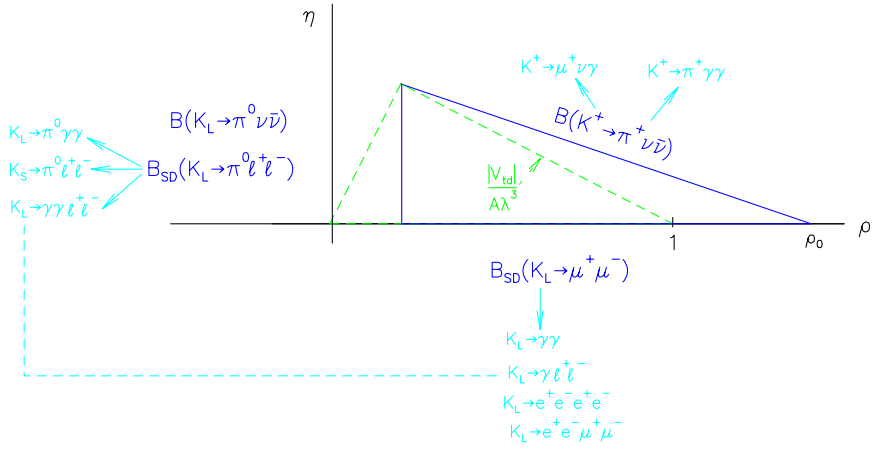}
  \caption{
$K$ decays and the unitarity plane.  The usual unitarity triangle
is dashed.  The triangle that can be constructed from rare $K$ decays
is solid.  See text for further details.
    \label{fig:triangle} }
\end{figure}

In the kaon sector experimental effort is now focussed on 
GIM-suppressed decays in which loops
containing weak bosons and heavy quarks dominate or at least
contribute measurably to the SM rate.  These 
processes include \kpnnp, \kmm,
\kpnn0, $K_L \to \pi^0 e^+ e^-$ and $K_L \to \pi^0\mu^+\mu^-$.
In the latter three cases the one-loop contributions violate CP. In \kpnn0 
this contribution completely dominates the decay\cite{Littenberg:1989ix}.  
Diagrams for such loops are shown in Fig.~\ref{fig:loops}.
Since the GIM-mechanism enhances the contribution of heavy quarks, in the 
SM these decays are sensitive to the product of couplings $V_{ts}^* V_{td}
\equiv \lambda_t$. Although it is perhaps most natural to write the branching 
ratio for these decays in terms of the real and imaginary parts of 
$\lambda_t$\cite{Littenberg:2000jn,Kettell:2002ep}, for comparison with 
results from the $B$ system
it is convenient to express them in terms of the Wolfenstein parameters,
$A$, $\rho$, and $\eta$.  Fig.~\ref{fig:triangle} shows the relation of
these 
rare kaon decays to the unitarity triangle.  The dashed triangle 
is the usual one derived from $V^*_{ub} V_{ud} + V^*_{cb} V_{cd} +
V^*_{tb} V_{td}  = 0$ ($\equiv \lambda_u + \lambda_c + \lambda_t$), 
the solid one indicates the information
available from rare kaon decays.   The
apex, $(\rho, \eta)$, can be determined from either triangle, and
disagreement between the $K$ and $B$ determinations implies physics beyond 
the SM. In Fig.~\ref{fig:triangle} 
the branching ratio closest to each side of the solid triangle can be used to
determine the length of that side.  The arrows leading outward from those
branching ratios point to processes that need to be studied either because
they could potentially constitute backgrounds, or because knowledge of them is
required to relate the one-loop process branching ratios to the lengths of
the triangle sides. $K_L \to \mu^+ \mu^-$, which can in principle determine 
the bottom of the triangle ($\rho$),
is the decay for which the experimental data is the best but for which
the theory is most problematical.  \kpnn0, which determines
the height of the triangle ($\eta$) is the cleanest theoretically, but for
this mode experiment falls far short of the SM-predicted 
level.  \kpnnp, which determines the hypotenuse, is nearly as clean as \kpnn0~
and has been observed (albeit with only three events).
Prospects for the latter are probably the best of all since it
is both clean and already within reach experimentally.

\subsection{\kpnnp}

From the point of view of theory \kpnnp~ is remarkably clean.
The often problematical hadronic matrix element can be calculated to 
$\sim$ 2\% via an isospin transformation from that of 
$K^+ \to \pi^0 e^+ \nu_e$\cite{Marciano:1996wy}.
The hard GIM suppression minimizes QCD corrections and
the long-distance contributions to this decay are very small.
A recent discussion of the latter with references to previous
work can be found in \cite{Isidori:2005xm}.

\kpnnp~ is very 
sensitive to $V_{td}$ - it is actually directly sensitive to the quantity
$\lambda_t$ as can be seen in Eq.~\ref{pnn}~\cite{Buchalla:1998ba}:

\begin{eqnarray}
B(K^+ \to \pi^+ \nu\bar\nu) = & \kappa_+ \bigl [ \bigl({\rm Im} \lambda_t X(x_t) \bigr)^2 +
\bigl (\lambda^4 {\rm Re} \lambda_c P_c(X) + {\rm Re} \lambda_t X(x_t) \bigr)^2 \bigr]
\label{pnn}
\end{eqnarray}
where $\lambda \equiv sin \theta_{Cabibbo}$, $x_t \equiv (m_t / m_W)^2$,
$X(x_t)$ and $P_c$ contain the top and
charm contributions respectively and will be discussed below, and
$\kappa_+$ is given by
\begin{eqnarray}
\kappa_+ \equiv &  r_{K^+} {3 \alpha^2 B(K^+ \to \pi^0 e^+ \nu_e) 
\over{2 \pi^2 sin^4 \theta_W \lambda^2} } = & (1.48 \pm 0.02) \times 10^{-4} \biggl [\frac{0.227}{\lambda} \biggr ]^2
\label{kappa}
\end{eqnarray}
The factor $3$ in  Eq.~\ref{kappa} comes from the summation over neutrino
flavors.  The appropriate values for $\alpha$ and $sin \theta_W$ to use in this
instance are discussed in \cite{Buras:2004uu}.  The isospin-breaking
parameter $r_{K^+} = 0.901$\cite{Marciano:1996wy}.

The Inami-Lim function\cite{Inami:1981fz}, $X(x_t)$ characterizing the
GIM suppression of the top contribution is also given in
\cite{Buras:2004uu}.  For the current measured value of $m_t$
\footnote{For this purpose the correct parameter is $m_t(m_t)$, which
is about 10 GeV lower than the pole mass.}, $X(x_t) = 1.464
\pm 0.041$ and it is proportional $m_t^{1.15}$ in this region. The QCD
correction to this function is $\leq$ 1\%.
\begin{eqnarray}
\lambda^4 P_c(X) =& \frac{2}{3} X^e_{NL} + \frac{1}{3} X^{\tau}_{NL}
\label{psubc}
\end{eqnarray}
gives the charm contribution, where the functions $X^l_{NL}$ are those
arising from the NLO
calculation~\cite{Buchalla:1994wq,Buchalla:1998ba}.  The QCD
correction leads to a $\sim$ 30\% reduction of the charm Inami-Lim function.
A recent assessment of the uncertainties in this 
calculation can be found in Buras {\it et al.}\cite{Buras:2004uu}. 
They give $\lambda^4 P_c(X) = 9.8 \times 10^{-4}$
with an 18\% error where the leading contributors are 12\% for the 
uncertainty in $\mu_c$, the scale parameter, and 8\% for the uncertainty in 
the charm quark mass, $m_c(m_c)$.  Very recently,  preliminary results of
a NNLO calculation were reported\cite{Buras:2005gr}, in which the former 
error was reduced to 5\% and the total to $\sim$10\%.  The resultant
branching ratio prediction can then be given as
\footnote{Ref~\cite{Isidori:2005xm} gives a correction to $P_c(X)$ due to
long-distance effects of $+0.04 \pm 0.02$ that is included in this
estimate.} :
\begin{eqnarray}
B(K^+ \to \pi^+ \nu\bar\nu) = & (8.0 \pm 0.4_{m_c} \pm 0.15_{\mu_c} \pm 0.9_{CKM} \pm 0.6_{m_t}) \times 10^{-11}
\label{brpnnp}
\end{eqnarray}
where the last two components, the parametric ones due to uncertainties in the 
CKM matrix elements and in $m_t$, will naturally be reduced as data in these 
areas is improved.  This would make a high precision measurement of 
$B(K^+ \to \pi^+ \nu\bar\nu)$ very interesting from a BSM point of view.

As mentioned above
the branching ratio can also be written in terms of the Wolfenstein
variables~\cite{Buras:1994ec} and one finds it is proportional
to $A^2 X^2(x_t) \frac{1}{\sigma} \bigl [ (\sigma \bar\eta)^2 + 
(\rho_0 - \bar\rho)^2 \bigr ] $ where $\sigma \equiv (1-\lambda/2)^{-2}$ and
$\rho_0 \equiv 1 + \frac{P_c(X)}{A^2 X(x_t)}$.
To a good approximation the amplitude is proportional to
the hypotenuse of the solid triangle in Fig.~\ref{fig:triangle}.  This is
equal to the vector sum of the line proportional to $V_{td}/A \lambda^3$
and that from $(1,0)$ to the point marked $\rho_0$. The length $\rho_0 -1$ 
along the real axis is proportional to the amplitude for the charm 
contribution to \kpnnp.  More precisely, $B(K^+ \to \pi^+ \nu\bar\nu)$ 
determines an ellipse of small eccentricity in the $(\bar\rho, \bar\eta)$ 
plane centered at $(\rho_0,0)$ with axes $r_0$ and $r_0/\sigma$ where 
\begin{eqnarray}
r_0 \equiv &
\frac{1}{A^2 X(x_t)} \sqrt{{\sigma B(K^+ \to \pi^+ \nu\bar\nu) \over{5.3 \times 10^{-11}}}}.
\label{r0}
\end{eqnarray}

\begin{figure}[h]
\centering
\includegraphics[angle=0, height=.45\textheight]{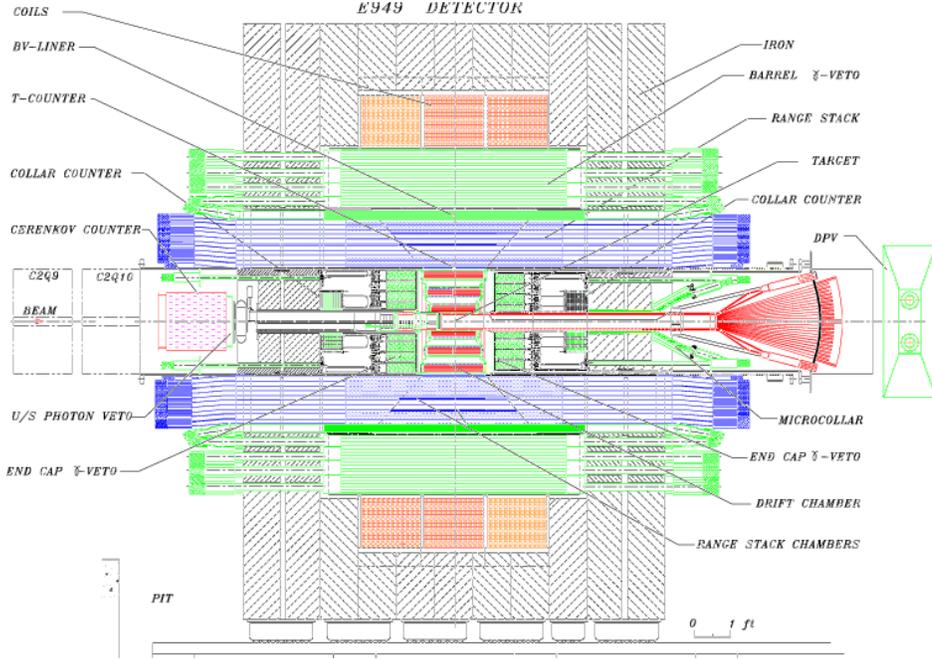}
  \caption{
E949 detector.
    \label{fig:det949} }
\end{figure}

	\kpnnp~ has been observed in the E787/949 series of experiments 
at the BNL AGS.
Like all previous experiments E787/949 used stopped $K^+$.  This gives 
direct access to the $K^+$ center of mass, and is conducive to hermetic 
vetoing.  The cylindrically symmetric detector, mounted inside a
1 Tesla solenoid~\cite{Atiya:1992vh}, is shown in Fig.~\ref{fig:det949}.
It sat at the end of a $\sim 700$ MeV/c positive beamline with two stages of
electrostatic separators.  This beamline provided an 80\% pure beam of 
$> 10^7~K^+$ per AGS cycle\cite{Doornbos:2000hb}.
Beam particles traversed a Cerenkov counter that identified $K^+$ and
$\pi^+$ and were tracked by two stations of MWPC's. They were
then slowed via $dE/dx$ in a BeO degrader followed by a Cu/scintillator shower
counter.
Approximately one quarter of the $K^+$ survived to exit the shower counter and
traverse a hodoscope before entering a scintillating fiber stopping
target.   A hodoscope surrounding the stopping target was used
demand a single charged particle leaving the target after a delay
of $\sim 0.12\, \tau_K$.  The emergent particle was tracked in a
low-mass cylindrical drift chamber with momentum resolution $\sim$1\%.
Additional trigger counters required the particle to exit the
chamber radially outward and enter a cylindrical array of scintillators 
and straw chambers, the ``Range Stack'' (RS), in which it was required
to stop in order for the event to be considered a $K^+ \to \pi^+ \nu\bar\nu$ 
candidate.  Each RS scintillation counter was read out by
phototubes on both ends, allowing a determination of the position
of hits along the beam direction via differential timing and pulse
height.  This facility, along with the pattern of counter pulse heights 
and the coordinates measured in two layers of straw chambers,
determined the range of the stopping particles to $\leq$3\%.  The detector 
design minimized ``dead'' material so that the kinetic energy could also be 
measured to $\sim$ 3\%.  Comparison of range, energy and momentum is a 
powerful discriminator of low energy particle identity. 
In addition, transient recorder readout of the RS
photomultipliers allowed the $\pi^+ \to \mu^+ \to e^+$ decay chain to
be used to identify $\pi^+$'s.  The combination of kinematic and
life-cycle techniques can distinguish pions from muons with a
misidentification rate of $\cal{O}$$(10^{-8})$.  Surrounding the RS
was a cylindrical lead-scintillator veto counter array and
adjacent to the ends of the drift chamber were endcap photon veto arrays
of undoped CsI modules\cite{Chiang:1995ar}.  There were also a number of
auxillary veto counters near the beamline as well as a veto in the beamline
downstream of the detector.

        Monte Carlo estimation of backgrounds was in general not
reliable since it was necessary to estimate rejection factors as high
as $10^{11}$ for decays occurring in the stopping
target.  Instead, methods to measure the background from
the data itself were developed, using the primary data stream as well as
data from special triggers taken simultaneously.  The principles adhered to
included:

\begin{itemize}
\item To eliminate bias, the signal acceptance region is kept hidden while
cuts are developed.

\item Cuts are developed on 1/3 of the data (evenly distributed throughout
the run) but residual background levels are measured only on the remaining 
2/3 after the cuts are frozen.

\item ``Bifurcated'' background calculation.  Background sources are identified
{\it a priori}.  Two independent high-rejection cuts are
developed for each background.  Each cut is reversed in turn as the other
is studied.  After optimization, the combined effect of the cuts can then be
calculated as a product.

\item Cuts are loosened in a controlled fashion to uncover correlations.  
If any are found, the correlated cuts are applied before the bifurcation 
instead of after it, and the background determination process repeated.

\item Signal and background functions are constructed and used in
likelihood analysis

\item Background calculations are verified though comparison with
data near the signal region.

\end{itemize}

In this way backgrounds can be reliably calculated at the $10^{-3}$
to $10^{-2}$ event level.

        All factors in the acceptance besides those of solid angle,
trigger and momentum interval were determined from data.

\begin{figure}[t]
\centering
\includegraphics[angle=0, height=.25\textheight]{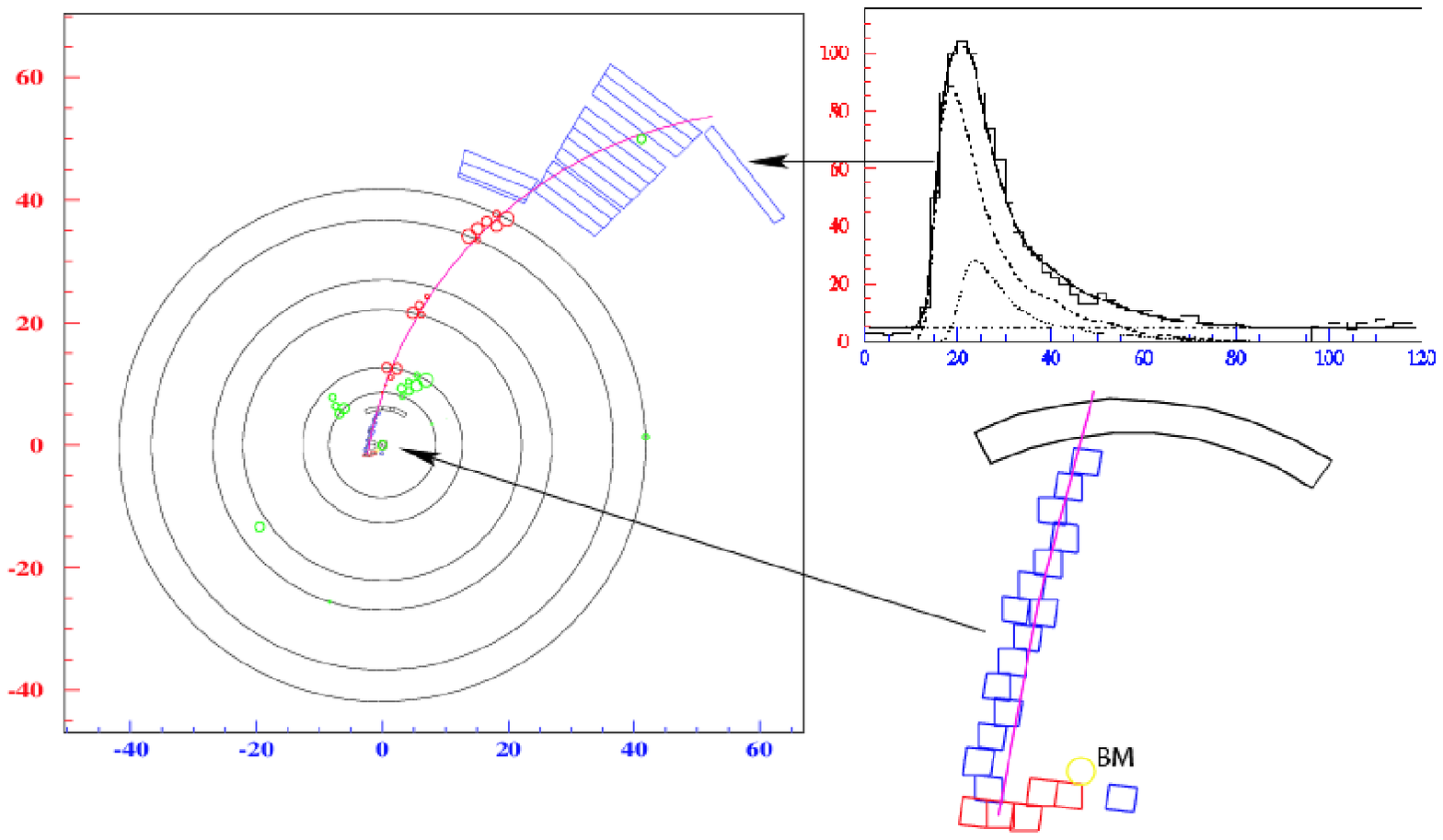}
\includegraphics[angle=0, height=.25\textheight]{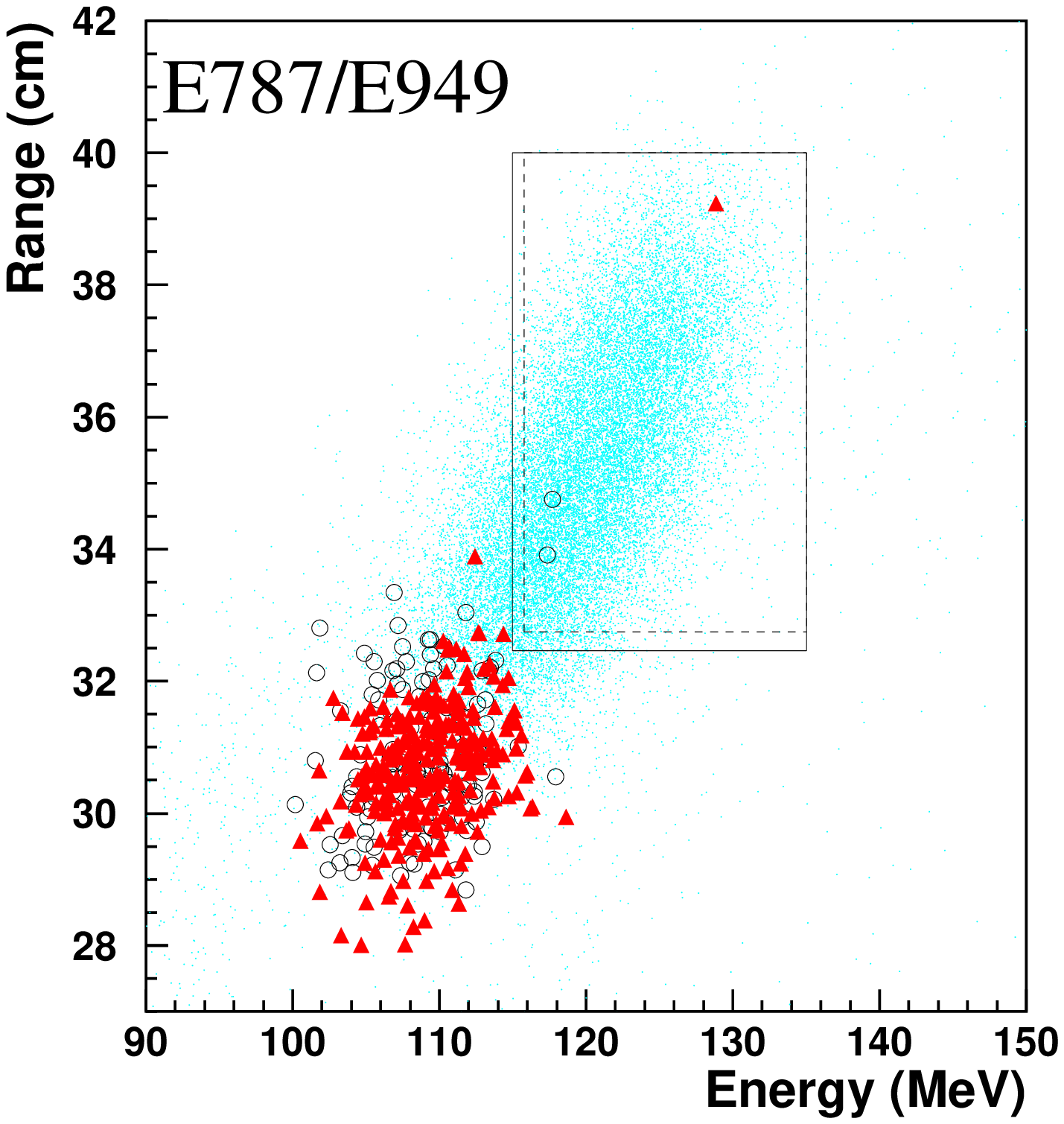}
  \caption{
{\bf Left:} E949 \kpnnp~event. {\bf Right:} Range vs energy of $\pi^+$ 
for E787/949 samples.  The circles are E787 data and the triangles E949 data.
The events 
around $E=108$ MeV are $K^+ \to \pi^+ \pi^0$ background.  The simulated
distribution of expected signal events is indicated by dots.
    \label{fig:e787} }
\end{figure}

	Using these techniques three \kpnnp~ events were observed,
two by E787~\cite{Adler:1997am,Adler:2001xv} and one by 
E949~\cite{Anisimovsky:2004hr}.  The combined result was a branching ratio 
\bkpnnp $= (1.47 {+1.30 \atop -0.89}) \times 10^{-10}$.  This is about 
twice as high as the prediction of  Eq.~\ref{brpnnp}, but
statistically compatible with it.

The total background to the two E787 events was measured to be 0.15 of an
event and that of the E949 event 0.3 of an event.
Thus E787/949 has developed methods to reduce the backgrounds
to a level sufficient to make a precise measurement of \kpnnp.

	It is possible to use the E787/949 results to extract information on
$\lambda_t$:
\begin{eqnarray}
0.24 < &|\lambda_t|/10^{-3} &   < 1.1 ~~~~ (68\% ~ CL) \\
-0.82< & Re(\lambda_t)/10^{-3}& < 1.1  ~~~~ (68\% ~ CL) \\
&Im(\lambda_t)/10^{-3}& < 1.1 ~~~~ (90\% ~ CL) 
\label{lt787}
\end{eqnarray}
These limits are not
competitive with what can be obtained using the full array of available 
phenomenological information, but they depend on fewer assumptions.
It will be very  interesting to compare the large value for $|V_{td}|$
suggested by the E787/949 result with the value that is extracted
from $\bar B_s - B_s$ mixing when it is finally observed.  

	From the first observation published in 1997, E787's results
for $B(K^+ \to \pi^+ \nu\bar\nu)$ have been rather high with respect
to the SM prediction.  Although there has never been a statistically 
significant disagreement with the latter, it has stimulated a
number of predictions in BSM theories.  
Table~\ref{tab:bsm787} lists a selection of such predictions.

	The \kpnnp~ data also yield an upper limit on
the process $K^+ \to \pi^+ X^0$ where $X^0$ is a massless weakly interacting
particle such as a familon~\cite{Wilczek:1982rv}. For E787 this was
$B(K^+ \to \pi^+ X^0)
<5.9 \times 10^{-11}$ at 90\% CL.  The case of $M_{X^0} > 0$ is discussed
below.

\begin{table}[b,h]
\caption{BSM predictions for $B(K^+ \to \pi^+ \nu\bar\nu)$} 
\begin{tabular}{|c|l|c|} \hline
\# & Theory & Ref. \\
\hline
1 & MSSM with no new sources of flavor- or CP-violation & \cite{Buras:2000dm} \\
2 & General MSSM & \cite{Buras:2004qb} \\
3 & Generic SUSY with minimal particle content & \cite{Buras:1999da} \\
4 & SUSY with non-universal A terms & \cite{Chen:2002eh} \\
5 & SUSY with broken R-parity & \cite{Bhattacharyya:1998be,Deandrea:2004ae} \\
6 & Upper limit from $Z'$ limit given by $K$ mass difference & \cite{Long:2001bc} \\
7 & Topcolor & \cite{Buchalla:1996dp} \\
8 & Topcolor-assisted Technicolor Model & \cite{Xiao:1999pt} \\
9 & Multiscale Walking Technicolor Model & \cite{Xiao:1999ps} \\
10 & $SU(2)_L \times SU(2)_R$ Higgs & \cite{Chanowitz:1999jj} \\
11 & Four generation model & \cite{Hattori:1999ap} \\
12 & Leptoquarks & \cite{Agashe:1996qm} \\
13 & L-R Model with right-handed Z' & \cite{He:2004it} \\
14 & Extension of SM to gauge theory of $J=0$ mesons & \cite{Machet:1999dj} \\
15 & Multi Higgs Multiplet Model & \cite{Grossman:1994jb} \\
16 & Light sgoldstinos & \cite{Gorbunov:2000cz}\\
17 & Universal extra dimensions & \cite{Buras:2002ej}\\
18 & 5-dimensional split fermions & \cite{Chang:2002ww} \\
19 & Randell-Sundrum scenario & \cite{Burdman:2002gr} \\
\hline
\end{tabular}
\label{tab:bsm787}
\end{table}

\begin{figure}[h]
\centering
\includegraphics[angle=0, height=.275\textheight]{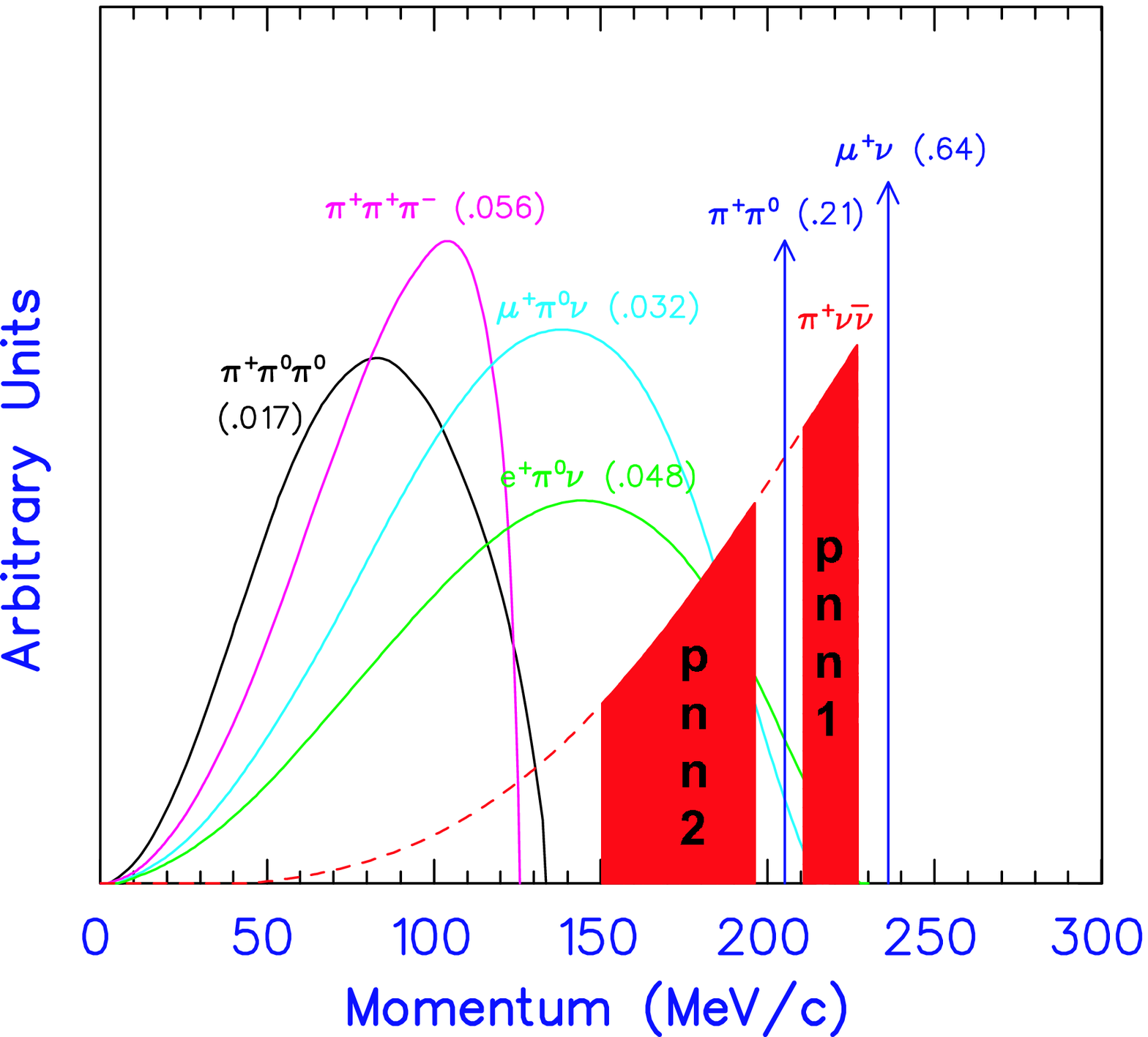}
\includegraphics[angle=0, height=.275\textheight]{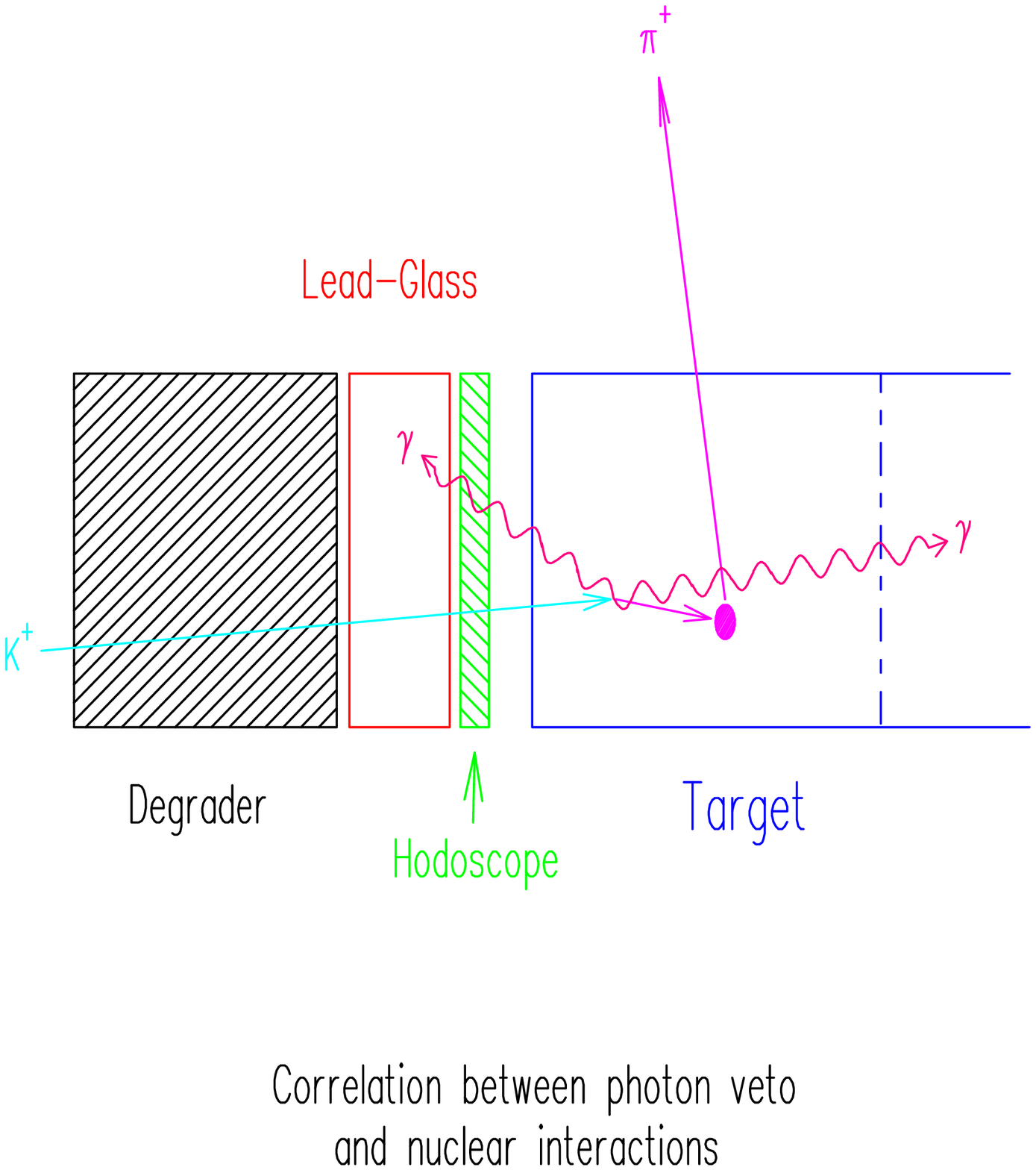}
  \caption{{\bf Left}: Center-of-mass momentum spectrum of $\pi^+$ from
$K^+ \to \pi^+ \nu\bar\nu$ compared with charged product spectra of the 
seven most common $K^+$ decays.  Filled areas indicate the portions
of the spectrum used in E787/949 analyses.  {\bf Right}: Cartoon of 
limiting background in the pnn2 region of the $K^+ \to \pi^+ \nu\bar\nu$
spectrum.  See text for details.
    \label{fig:pnn2} }
\end{figure}

	Fig.~\ref{fig:pnn2} left shows the $\pi^+$ momentum spectrum
from $K^+ \to \pi^+ \nu\bar\nu$ in the SM, along with the charged
track spectra from other kaon decays.

\begin{figure}[h]
\centering
\includegraphics[angle=0, height=.300\textheight]{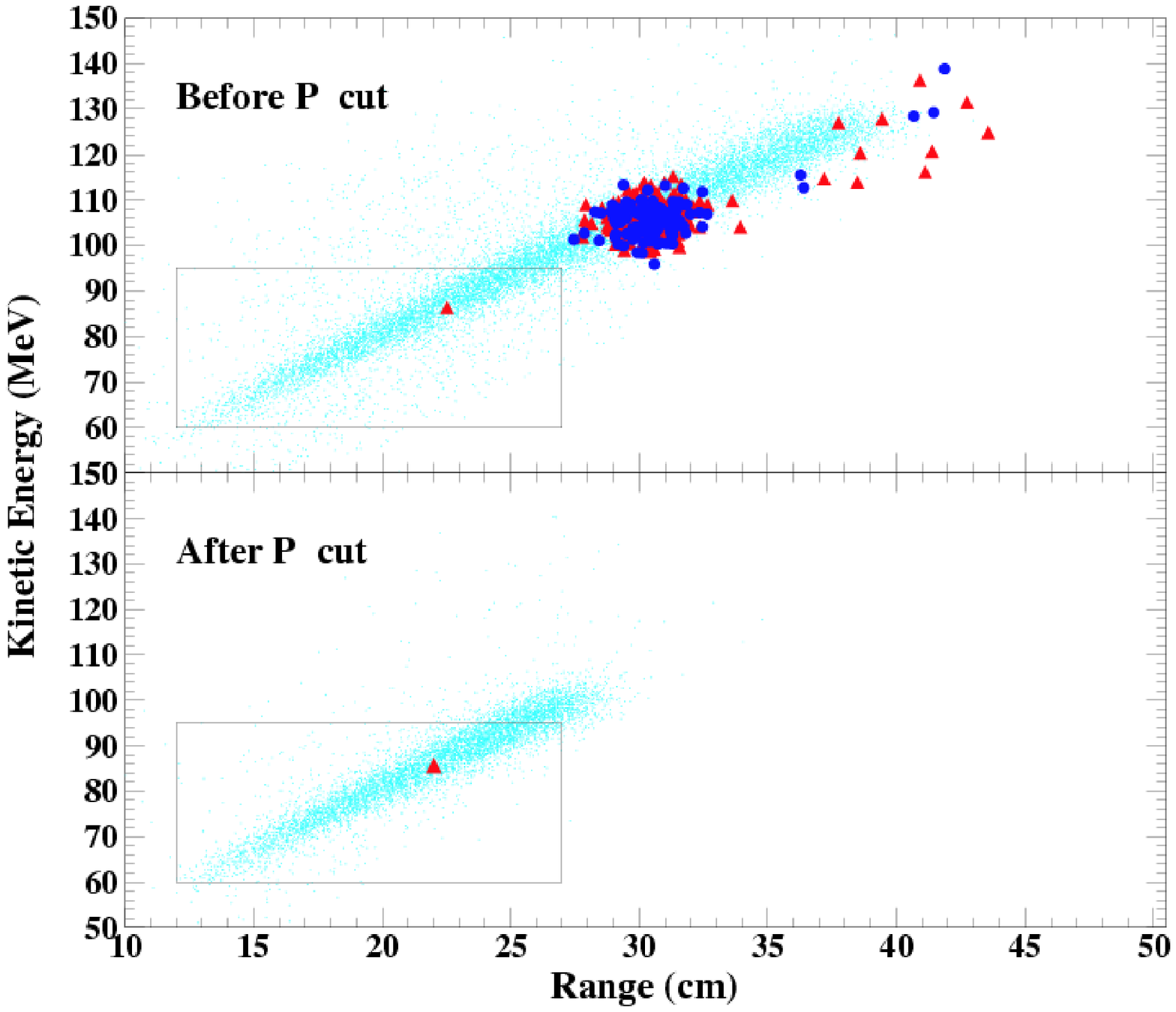}
\includegraphics[angle=0, height=.300\textheight]{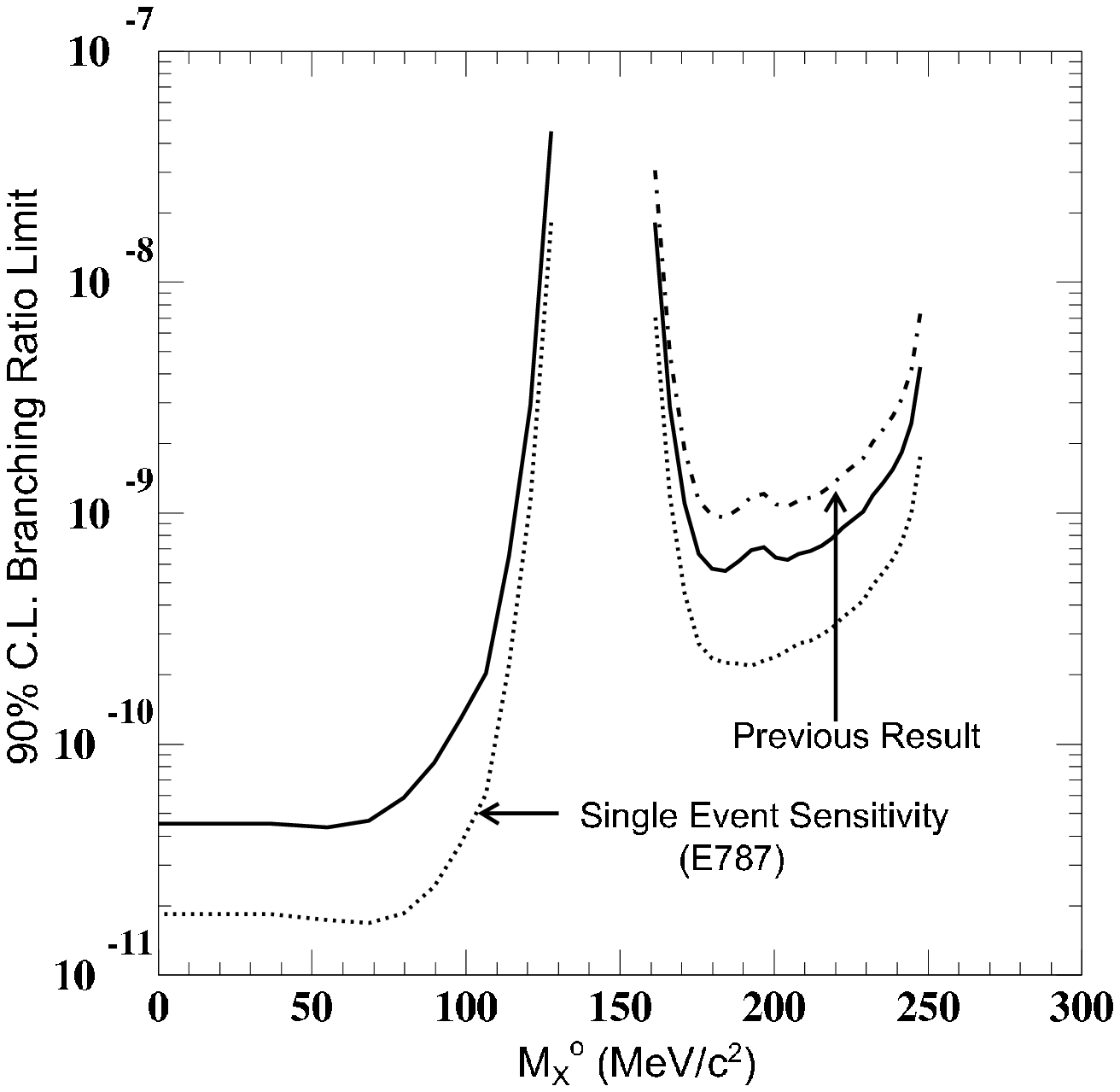}
  \caption{{\bf Left}: Signal plane for $K^+ \to \pi^+ \nu\bar\nu$
analysis of soft $\pi^+$ region. 
{\bf Right}: Limit on $K^+ \to \pi^+ X^0$ vs $m_X$.  Vertical lines
indicate location of events.  For comparison previous limits are
indicated as is the single event sensitivity limit of E787.
    \label{fig:pnn2_rslt} }
\end{figure}

	E787/949 is sensitive to the filled-in regions of the $K^+ \to \pi^+
\nu\bar\nu$ spectrum in Fig.~\ref{fig:pnn2}.  However all the results
mentioned so far come from the region on the right (``pnn1''), in which the
momentum of the $\pi^+$ is greater than than of the $\pi^+$ from
$K^+ \to \pi^+ \pi^0$ (205 MeV/c) .  The region on the left (``pnn2'') 
contains a larger portion of 
the signal phase space, but is more vulnerable to background from 
$K^+ \to \pi^+ \pi^0$.  It is relatively easy for the $\pi^+$ to lose energy 
through nuclear interactions.  Moreover, there is an unfortunate
correlation between nuclear scattering in the stopping target and the 
relatively weaker
photon veto in the beam region.  Fig.~\ref{fig:pnn2} right 
illustrates this problem.  A $K^+$ decays with the $\pi^+$ pointing downstream
(the $\pi^0$ must then be pointing upstream).  Normally such a decay
would not trigger the detector, but here the $\pi^+$ undergoes a 90$^\circ$
scatter, loses enough momentum to get into the accepted range
and heads for the drift chamber.  At the same time the $\pi^0$ decays
asymmetrically, with the high energy photon heading upstream, to where
the veto is least capable and the low energy photon heading downstream,
toward another relatively weak veto region.  This sequence of events is 
unlikely, but 20\% of $K^+$ decay to $\pi^+ \pi^0$, and one is trying to
study a process that happens one in ten billion times.  The fact that the
same scatter both down-shifts the $\pi^+$ momentum and aims the $\pi^0$
at the weak veto region confounds the usual product of rejection factors
so effective in the pnn1 region.  A test analysis using 1996 \& 1997
data was undertaken to determine whether methods could be developed
to overcome this background.  These leaned
heavily on exploiting the transient digitized signals from the stopping
target target scintillating fibers.  At the cost of giving up some
acceptance (using only decays later than 0.5 $\tau_K$) one could
detect evidence of $\pi^+$ scattering occluded by kaon signals
in the critical target elements.  In this way a single event sensitivity
of $7 \times 10^{-10}$ was achieved with a calculated background of $1.22$
events.  Fig.~\ref{fig:pnn2_rslt} left shows the resulting distribution
of $\pi^+$ kinetic energy and range for surviving candidates.
The top left shows the distribution before the final cut on $\pi^+$
momentum.  The peak at $T_{\pi} \sim 108$ MeV, $R_{\pi} \sim 30.5$ cm
is due to $K^+ \to \pi^+ \pi^0$ events.  After the final cut, one
event remains, consistent with the background estimation.  This yields
$B(K^+ \to \pi^+ \nu\bar\nu) < 2.2 \times 10^{-9}$ at 90\% CL
~\cite{Anisimovsky:2004hr}, consistent
with other E787/949 results.  This kinematic region is particularly sensitive
to possible BSM effects which produce scalar or tensor pion spectra
(rather than the vector spectrum given by the SM).  One can combine
this region with the high momentum region to get 90\% CL upper limits of 
 $2.7 \times 10^{-9}$ and $1.8 \times 10^{-9}$ for scalar and tensor 
interactions,  
respectively.  These measurements are also sensitive to $K^+ \to \pi^+
X^0$ where $X^0$ is a hypothetical stable weakly interacting particle
or system of particles.  Fig.~\ref{fig:pnn2_rslt} right shows 90\% CL upper
limits on $B(K^+ \to \pi^+ X^)$ together with the previous limit from
\cite{Atiya:1993qr}.  The dotted line in the figure is the single
event sensitivity.  Note that progress in the sensitivity of this kind of 
search is starting to be impeded by ``background'' from \kpnnp .

E949~\cite{e949}, which ran in 2002 was based on an upgrade of the E787
detector.  It was
improved in a number of ways with respect to E787: thicker and more
complete veto coverage, augmented beam instrumentation, higher
capacity DAQ, more efficient trigger counters, upgraded chamber
electronics, auxiliary gain monitoring systems, etc.  In addition SM
sensitivity was anticipated for the pnn2 region ($140 <p_{\pi^+} <
190$ MeV/c).  Several of the upgrades were aimed at exploiting this
region, and based on the test analysis discussed above, a signal/background
of 1:1 is expected for this part of the spectrum.
Using the entire flux of the AGS for 6000 hours, E949 was
designed to reach a sensitivity of $\sim 10^{-11}$/event. 
In 2002 the detector operated well at fluxes nearly twice as 
high as those typical of E787, but unfortunately DOE support of the 
experiment was terminated after that first run.  Aside from the 
results of the analysis of the pnn2 region that is still continuing, further
progress in \kpnnp~ will have to come from experiments yet to be
mounted.

There are currently two initiatives for future \kpnnp~ experiments.
One is a J-PARC LOI~\cite{jparc4} for a higher-sensitivity stopping
experiment.  This is very like E787/949 in conception, but with many
improvements in detail.  These include a lower incident beam momentum
(leading to a higher stopping efficiency and a better signal/random
rate ratio), higher B-field (leading to better momentum resolution and
a more compact geometry), higher granularity (leading to greater rate
capability and muon rejection power), brighter scintillators, crystal
photon vetoes, etc.  The goal of this experiment is to observe 50
events at the SM-predicted level.  The possible schedule for such an
experiment may become clear later this year when full proposals for
J-PARC experiments are expected to be requested.  A lower limit may be
deduced from the fact that the first hadron beams are scheduled to be
available sometime in 2008.

Although the stopped-$K^+$ technique is now well-understood, and one could
be reasonably sure of the outcome of any new experiment of this type, to 
get really large samples of \kpnnp~ ($\ge$100 events), it will almost
certainly be necessary to go to an in-flight configuration.  There have
been a series of attempts to initiate such an experiment, most recently the
P326 proposal to CERN~\cite{p326}.  This experiment exploits
newly developed tracking technology to allow the use of an extremely
intense ($\sim$ 1 GHz) unseparated 75 GeV/c beam.  Charged beams of this
intensity have been used to search for good-signature kaon decays 
such as $K^+ \to \pi^+ \mu^+ e^-$~\cite{Appel:2002xz}, but P326 is a
departure for a poor-signature decay for which high-efficiency vetoing is 
required.  

\begin{figure}[h]
\centering
\includegraphics[angle=0, height=.375\textheight]{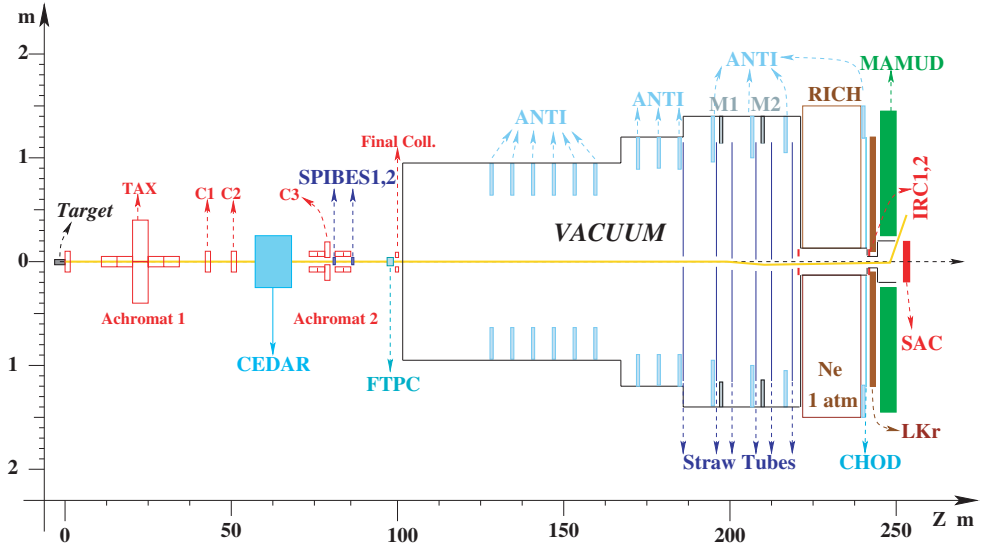}
  \caption{
P326 detector for $K^+ \to \pi^+ \nu\bar\nu$.
    \label{fig:p326} }
\end{figure}

Fig.~\ref{fig:p326} shows the layout of the proposed experiment.  Protons
from the 400 GeV/c SPS will impinge on a 40cm Be target.  Positive secondaries 
with 
momenta within $\pm$2\% of 75 GeV/c will be taken off in the forward direction.
The $\sim$5\% of $K^+$ in the beam will be tagged by a differential Cerenkov 
counter (CEDAR).  The 3-momenta of all tracks will be measured in 
a beam spectrometer with three sets of ``GIGATRACKER'' detectors
(fast Si micro-pixels \& micro-mega TPCs).  The expected performance is
$\sigma_p =$ 0.3\%, $\sigma_{\theta} = 10 \mu$r and $\sigma_t=$ 150ps.
The beamline has been carefully designed to hold
the muon halo that contributes to detector random rates to
the order of 10MHz.  The beam will continue through the apparatus in
vacuum.  About 10\% of the $K^+$ will decay in a 60m-long fiducial region.
The remainder of the beam will be conducted in vacuum out of the detector 
region.  Photons from $K^+$ decays will be  
detected in a series of ring vetoes (at wide angles), by an
upgraded version of the NA48
liquid Krypton calorimeter (at intermediate angles) and
by two dedicated inner veto systems.  Charged decay tracks 
will be momentum-analyzed in a two-dipole straw-tube spectrometer
($<$1\% resolution on pion momentum and 50-60 $\mu$r resolution
on $\theta_{K \pi}$ are necessary).  Downstream
of the spectrometer a RICH filled with Ne at 1 atm will help distinguish
signal pions from background muons. This is to be followed by a charged
particle hodoscope of multigap glass RPC design (100 ps resolution is
required).  Behind the hodoscope is the ``MAMUD'' muon veto, a magnetized
iron-scintillator sandwich device to complete the pion/muon distinction.
Its 5T-m bending power serves to kick the beam out of the way of the small
angle photon veto at the back of the detector.
\begin{figure}[h]
\centering
\includegraphics[angle=0, height=.35\textheight]{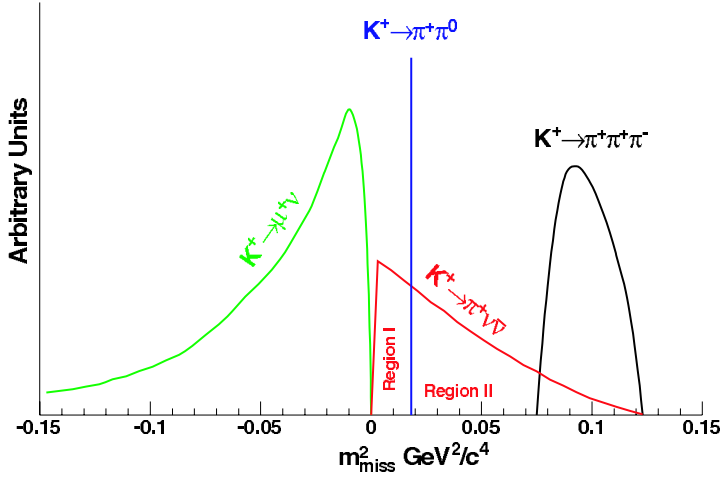}
  \caption{
Missing mass squared under the pion assumption for P326 (see text).
    \label{fig:k326} }
\end{figure}
Fig.~\ref{fig:k326} shows the distribution in the square  of missing mass 
recoiling from the assumed
$\pi^+$ expected in this experiment \footnote{The $K^+ \to \mu^+ \nu$
peak is shifted and smeared out by this assumption.} (this plot corresponds
to Fig.~\ref{fig:pnn2} Left for E787/949).  A resolution of $1.1 \times
10^{-3}\,$GeV$^2$/c$^4$ in this quantity is required to get sufficient
signal/background.  

The collaboration proposes to build this experiment in time to
begin taking data in 2009.  A two-year would accumulate $\sim$80 events
with a 8:1 signal to background.

\subsection{\kpnn0}
\kpnn0 is generally considered to be the most attractive target in the kaon 
system, since 
\begin{enumerate}
\item it is
direct CP-violating to a very good
approximation~\cite{Littenberg:1989ix,Buchalla:1998ux} (in the SM \bkpnn0 
$\propto\eta^2$) and  
\item the rate can be rather precisely calculated in the SM or almost any
extension thereof~\cite{Bryman:2005xp}.
\end{enumerate}
Like \kpnnp\ the
hadronic matrix element can be obtained from $K_{e3}$, but
unlike \kpnnp, it has no significant contribution from charm.  Consequently, the
intrinsic theoretical uncertainty connecting  \bkpnn0\ to
the fundamental short-distance parameters is only about 2\%.  
Note also that in the SM \bkpnn0~
is directly proportional to the square of $Im \lambda_t$ and
that $Im \lambda_t = - $$\cal{J}$$/[\lambda (1-\frac{\lambda^2}{2})]$
where $\cal{J}$ is the Jarlskog invariant~\cite{Jarlskog:1985ht}.
Thus a measurement of \bkpnn0~determines the area of the unitarity
triangles with a precision twice as good as that on \bkpnn0~itself.

        \bkpnn0\ can be bounded indirectly by measurements of \bkpnnp\
through a nearly model-independent relationship pointed out by
Grossman and Nir~\cite{Grossman:1997sk}.  The application of this to
the E787/949 results yields \bkpnn0$<1.4 \times 10^{-9}$ at 90\% CL.
This is far tighter than any extant direct experimental limit.
To actually observe \kpnn0 at the SM level ($\sim 3 \times 10^{-11}$),
one will need to improve on the current state of the art  by some four 
orders of magnitude.

\begin{figure}[ht]
\centering
 \includegraphics[angle=0, height=.3\textheight]{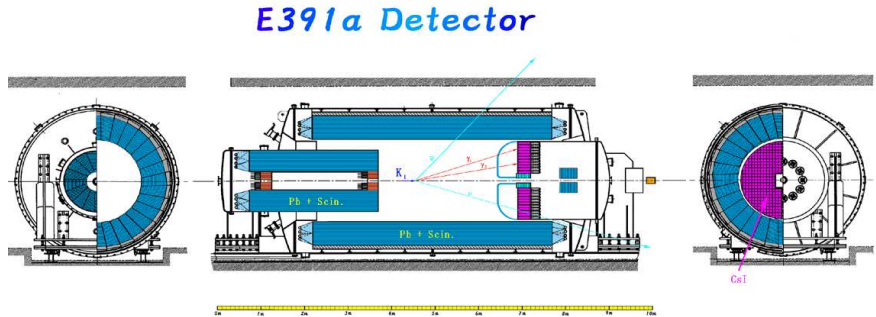}
  \caption{KEK E391a detector for $K_L \to \pi^0\nu\bar\nu$
    \label{fig:e391a} }
\end{figure}

The first dedicated \kpnn0~ experiment,
KEK E391a~\cite{Inagaki:1997gc}, mounted at the KEK 12
GeV proton synchrotron, aims to achieve sensitivity comparable to the
indirect limit.  Although it will not approach the SM level, it will
serve as a test for a future more sensitive experiment to be performed
at J-PARC~\cite{jparc5}.  E391a features a
carefully designed ``pencil'' beam~\cite{Watanabe:2005gc}
with average $K_L$ momentum $\sim 2$ GeV/c. Fig.~\ref{fig:e391a} shows a
layout of the detector.

A particular challenge of the E391a approach is to achieve extremely
low photon veto inefficiency.  The photon veto system consists of two
cylinders.  The inner, more upstream barrel is intended to suppress
beam halo and reduce confusion from upstream $K_L$ decays.  Roughly
4\% of the $K_L$'s decay in the 2.4m fiducial region between the end
of the inner cylinder and the charged particle veto in front of the
photon detector.  Signal photons are detected in a multi-element
CsI-pure crystal calorimeter~\cite{Doroshenko:2005gd}.  The entire
apparatus operates in vacuum.  Physics running began in February 2004.
A second run took place early this year, and a third is scheduled
for the fall.

In this experiment events with two showers in the calorimeter and no
additional activity are examined to determine whether any point along
the fiducial section of the beamline results in a reconstructed mass
consistent with a $\pi^0$.  If so, the $p_T$ can then be determined.
Cuts are imposed on the shower patterns and energies, the $Z_V$ and
the $p_T$.  In addition, events consistent with $\eta \to \gamma \gamma$
were discarded.

\begin{figure}[ht]
\centering
\includegraphics[angle=0, height=.3\textheight]{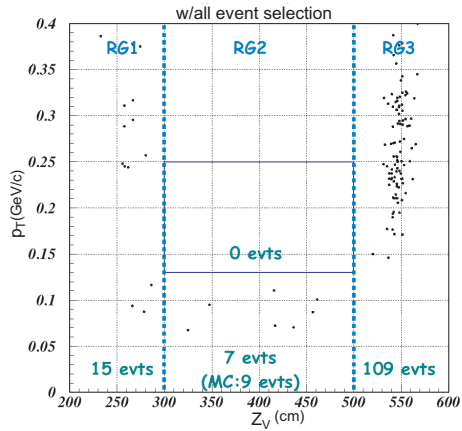}
  \caption{$p_T$ vs $Z_V$ E391a events passing all other cuts 
(from first 10\% of Run 1).  Signal region is the central rectangle.
    \label{fig:r391a} }
\end{figure}

	Fig.~\ref{fig:r391a} shows the distribution of residual
candidates in $p_T$ vs $Z_V$, when all other cuts are applied for the
first 10\% of Run 1.  No events were observed
in the signal region, with an expected background of 0.03$\pm$0.01 event
\footnote{Not all backgrounds have yet been calculated.}.
With $1.14 \times 10^9 \, K_L$ decays and an acceptance of 0.73\%,
they obtained a preliminary result of $B(K_L \to \pi^0 \nu\bar\nu)< 2.86
\times 10^{-7}$ at 90\% CL~\cite{Sakashita:2005}.

\begin{figure}[ht]
\centering
 \includegraphics[angle=0, height=.275\textheight]{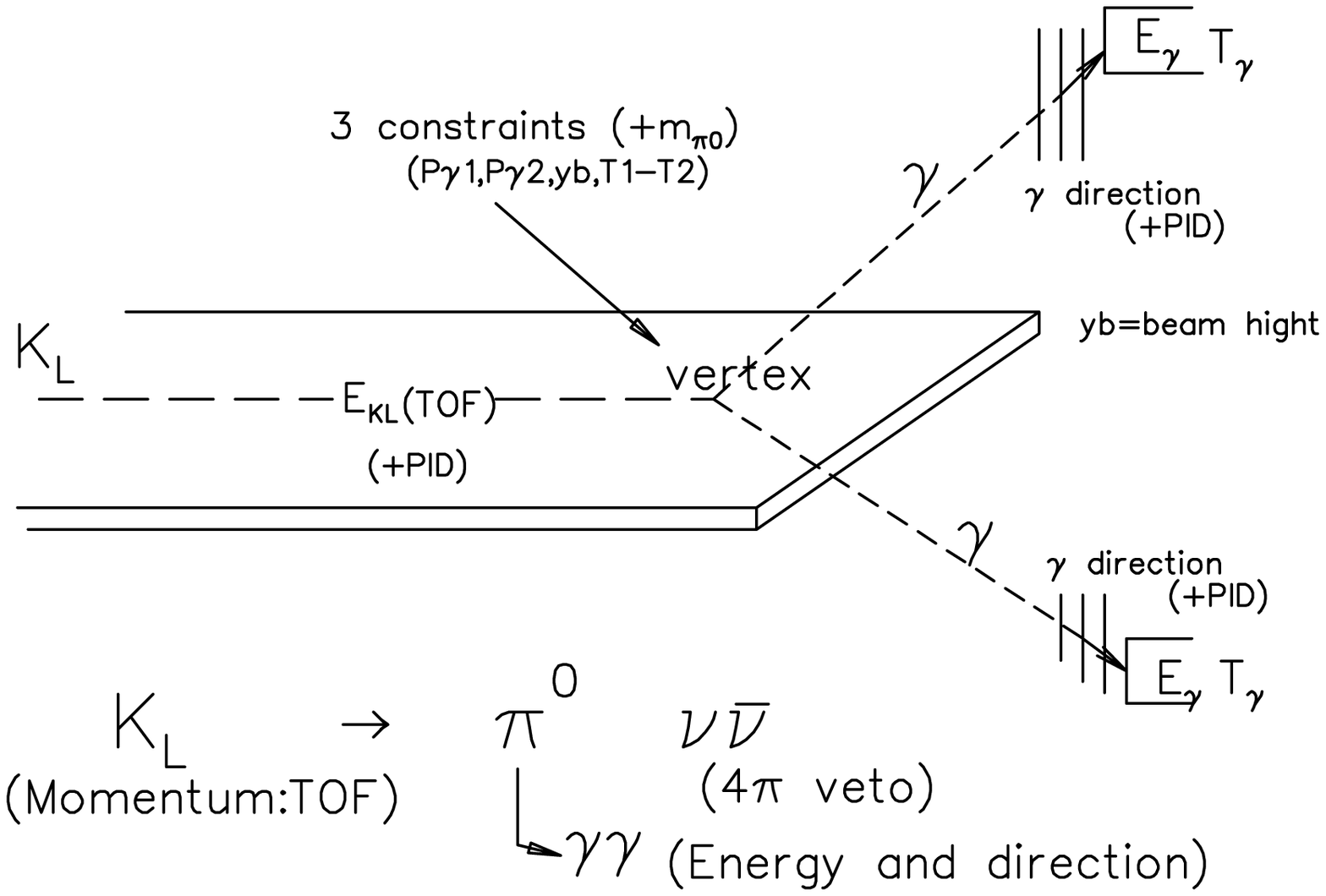}
  \caption{Principles of KOPIO $K_L \to \pi^0\nu\bar\nu$ experiment
    \label{fig:k_prin} }
\end{figure}

        A different approach is taken by the KOPIO
experiment\cite{Bryman:2001hs} (E926) at BNL, exploiting the intensity
and flexibility of the AGS to make a high-flux, low-energy,
microbunched $K_L$ beam.  The principles of the experiment are
illustrated in Fig.~\ref{fig:k_prin}.

The AGS proton beam will be microbunched at 25 MHz by imposing
upon it a train of empty RF buckets as it is extracted from the 
machine\cite{Glenn:1997dn}.  The AGS's $10^{14}$ protons will be
spilled out over $\sim$4.7s.
A neutral beam will be extracted at 
42.5$^{\rm o}$ to obtain a spectrum soft enough
to permit time-of-flight determination of the $K_L$ velocity.
The beam contains about $3 \times 10^8 \, K_L$ and 
$5 \times 10^{10}$ neutrons per cycle.
The large production angle also softens the neutron spectrum 
so that they (and the $K_L$) are by and large below threshold for the
hadro-production of $\pi^0$'s.  The beam region will be evacuated to
$10^{-7}$ Torr to further minimize such production.  With a 10m beam channel 
and this low energy beam, the contribution of hyperons and $K_S$ 
to the background will
be negligible.  The profile of the beam is ribbon-like (4mr $\times$ 90mr)
to facilitate 
collimation of the large aperture and to provide an extra constraint for 
reconstruction of the decay vertex.  All possible quantities are measured:
the $K_L$ momentum, the final state photon angles as well as energies
and times.  Effective kinematic rejection of background is 
then possible.

\begin{figure}[ht]
\centering
 \includegraphics[angle=0, height=.30\textheight]{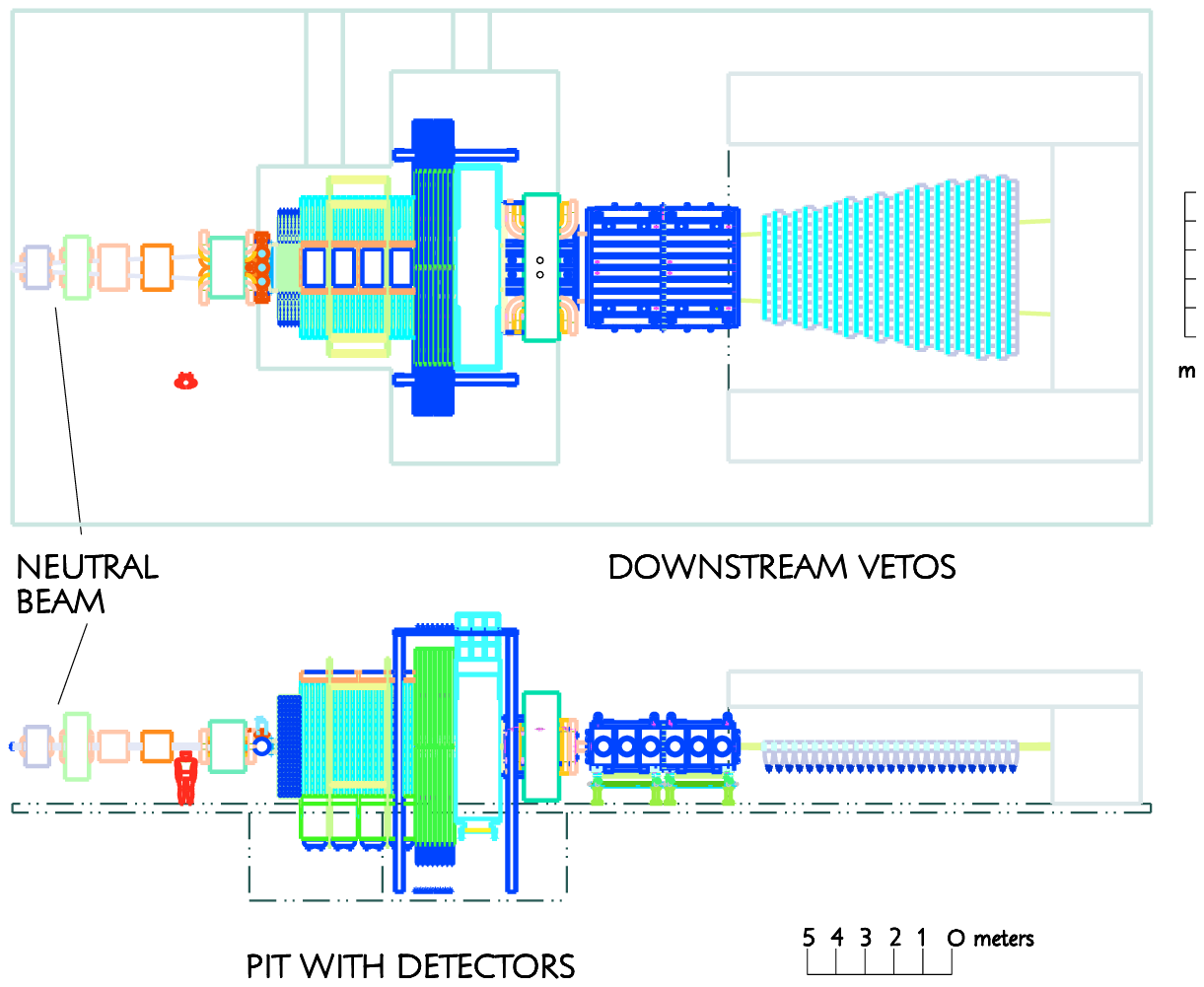}
  \caption{Layout of the KOPIO detector.
    \label{fig:kopiod} }
\end{figure}

\begin{figure}[ht]
\centering
 \includegraphics[angle=0, height=.25\textheight]{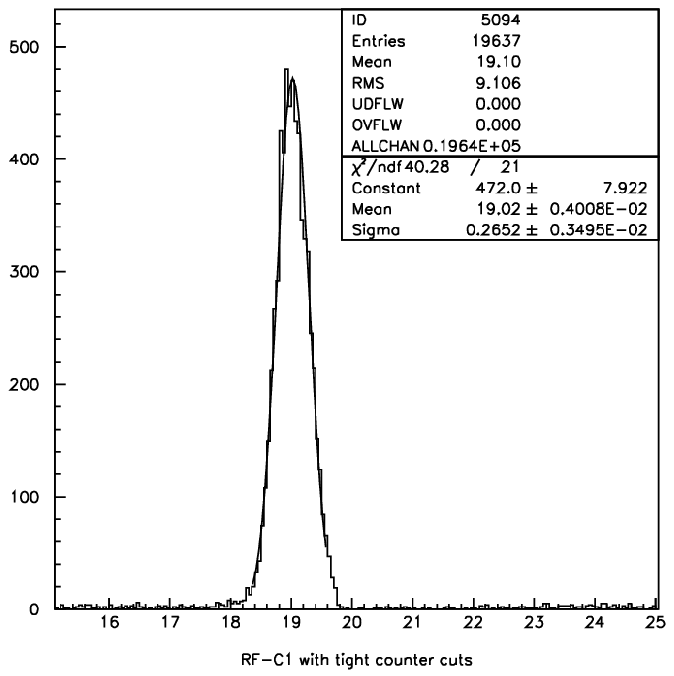}
 \includegraphics[angle=0, height=.25\textheight]{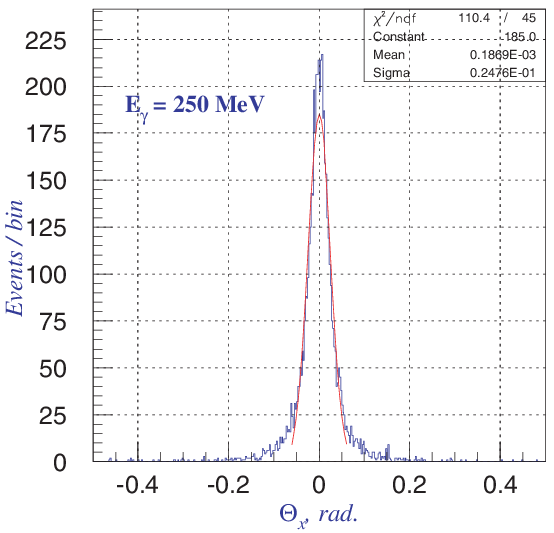}
  \caption{Tests of KOPIO components.  {\bf Left:} demonstration of 
microbunching of the AGS proton beam (93 MHz RF cavity at 22 KV).  
{\bf Right:} angular resolution of prototype preradiator for 250
MeV photons (in a tagged photon beam).
    \label{fig:k_tests} }
\end{figure}

The layout of the  experiment is shown in Fig.~\ref{fig:kopiod}.  
$K_L$ decays from a $\sim 3$m fiducial region will be accepted ($\sim$
8\% of the $K_L$ decay in this region).  Signal 
photons impinge on a 2 $X_0$ thick preradiator capable of measuring their 
direction to $\sim 25$mrad.  An alternating drift chamber/scintillator plane 
structure will also allow good measurement of the energy deposited in 
the preradiator.  A high-precision shashlyk calorimeter downstream of 
the preradiator will complete the energy measurement.  The photon
directions allow the decay vertex position to be determined.
This can be required to lie within the beam envelope, eliminating many
potential sources of background.  
Combined with the target position and time of flight information,
the vertex information provides a measurement of the $K_L$ 3-momentum so that
kinematic constraints as well as photon vetoing are available to suppress
backgrounds.  The leading expected background is \kp0, initially
eight orders of magnitude larger than the SM-predicted signal.  
However since $\pi^0$'s from this background have characteristic 
signatures in the $K_L$ center of mass, effective kinematic cuts can be
applied.  Similar remarks pertain to almost all expected backgrounds.
This reduces the burden on the photon veto system
surrounding the decay region to the point where the hermetic 
veto techniques proven in E787/949 are sufficient.  In fact most of
the techniques necessary for KOPIO have been proven in previous 
experiments or in prototype tests.  Fig.~\ref{fig:k_tests} shows results on
two of the more critical aspects of the experiment.

	On the left of the figure is the result of a test of
beam microbunching, showing an rms of 265 psec.  This is sufficient
for KOPIO's purposes, although it still needs to be demonstrated at a
25 MHz repetition rate.  Other beam tests have demonstrated the required
$\le 0.001$ extraction of protons between the microbunches.
On the right is a plot of photon angular
resolution obtained with a 6-plane prototype of the preradiator.  A
tagged beam at the National Synchrotron Light Source (NSLS) provided the
photons.  A resolution of $25 \,$mrad is observed for 250 MeV photons, in
line with GEANT simulation.  This resolution is sufficient for KOPIO.

	The electromagnetic calorimeter following the preradiator will
be a 5m $\times$ 5m array of high resolution shashlyk modules.  The
required resolution of 3\%/$\sqrt(E)$ has been demonstrated in prototypes
tested in the NSLS tagged photon beam~\cite{Atoian:2003qb,Atoian:2004}.  
The combination of preradiator
and calorimeter will measure photon energies to $\sim$2.7\%$/\sqrt{E}$.

	The upstream veto wall will be 18 r.l. thick lead-scintillator
shower counters read out via wavelength-shifting fibers~\cite{Yershov:2004yd}. 
The cylindrical barrel will be shashlyk of similar structure to those of
the calorimeter, but larger transversely and truncated to fit into a
cylinder. The demands on the performance of these counters are
comparable to that demonstrated in the E787/949 barrel veto that has
similar structure.  It is also planned to use the barrel shashlyk for
positive photon detection.  There are a series of downstream vetoes
abutting the beam, using plate counter technology similar to that of
the upstream wall.  Finally
it will also be necessary to veto within the beam, which is
very challenging but which is facilitated by the low average energy of the
beam neutrons.  This will be accomplished by a series of lead-aerogel
shower counters (the ``catcher'' veto).  For the most part charged
particles created by the neutrons are below the Cerenkov threshold of
the aerogel and are so invisible to these counters.  

Another very important element is charged particle vetoing 
needed to eliminate backgrounds such as $K_L \to \pi^0 \pi^+ \pi^-$.
A very high performance system will be mounted in the decay region
vacuum and at the margins of the downstream beam pipe.  Behind the calorimeter
will be a dipole magnet with field oriented to sweep charged particles
traveling in the beam direction upwards or downwards into veto counters 
outside the beam profile.
	
KOPIO aims
to collect about $150$ \kpnn0\ events with varying signal to background.
This will permit $\eta$ to be determined to $\sim 7\%$,
given expected progress in measuring $m_t$ and $V_{cb}$.  KOPIO will
run during the $\sim$20 hours/day the AGS is not needed for injection 
into RHIC.  The experiment is presently awaiting funding
\footnote{On August 11, 2005, RSVP, of which this experiment is a component,
was canceled by the US National Science Foundation}. 

\subsection{$K_L \to \mu^+ \mu^-$}

\begin{figure}[h]
\centering
 \includegraphics[angle=0, height=.20\textheight]{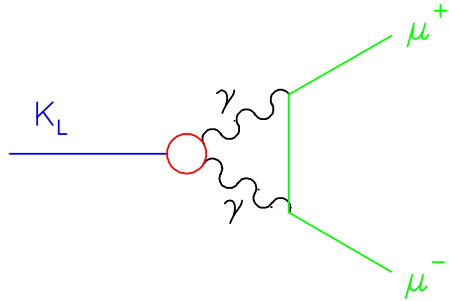}
  \caption{
Long distance contribution to  $K_L \to \mu^+ \mu^-$.
    \label{fig:mulong} }
\end{figure}

The short distance component of this decay, which arises out of the
diagrams shown in Fig.~\ref{fig:loops}, can be  quite reliably
calculated in the SM\cite{Buchalla:1994wq}.  The most recent
measurement of its branching ratio\cite{Ambrose:2000gj} based on $\sim 6200$
events gave $B(K_L \to \mu^+ \mu^-) = (7.18 \pm 0.17)\times 10^{-9}$.  
However the actual measurement was the ratio $R_{\mu\mu} \equiv 
B(K_L \to \mu^+ \mu^-)/B(K_L \to \pi^+ \pi^-)$, and the value of 
the denominator has recently changed significantly, so 
that the current value of $B(K_L \to \mu^+ \mu^-)$ is $(6.87 \pm 0.12) \times 
10^{-9}$\cite{Eidelman:2004wy}\cite{PDBook05} (this average also includes the results
of older, less precise experiments).  This is quite a precise determination
for such a rare decay.
However \kmm~ is dominated by long distance effects, the
largest of which, the absorptive contribution mediated by $K_L \to
\gamma\gamma$ shown in Fig.~\ref{fig:mulong}, accounts for 
$(6.64 \pm 0.07)\times 10^{-9}$.

Subtracting the two, yields $(0.23 \pm 0.14) \times 10^{-9}$ for 
dispersive part of \bkmm~.  One can do
a little better than this in the following way.  As mentioned
above, the actual quantity
measured in Ref~\cite{Ambrose:2000gj} was $R_{\mu\mu} =
(3.48 \pm 0.05)\times 10^{-6}$.  It is necessary to subtract from this
measured quantity the ratio $B^{abs}_{\gamma\gamma}(K_L \to
\mu^+\mu^-)/ \bkppr$.  Eq.~\ref{foo} shows the components of
this latter ratio, obtained from Ref.~\cite{Hagiwara:2002pw}
augmented by a recent measurement 
$\Gamma(K_S \to \pi^+ \pi^-)/\Gamma(K_S \to \pi^0 \pi^0)= 2.236 \pm 0.003
\pm 0.015$~\cite{Aloisio:2002bs}, whose product is $(3.344 \pm 0.053)
\times 10^{-6}$.
\newpage
\bea
{\rm recently~measured}~~~~~~~~~~~~~~~~~~~~ \cr
{\rm by~KLOE}~~~~~~~~~~~~~~~~~~~~~~~~~ \cr
{\rm last~measured~by}~~~~~~~~~~~|~~~~~~~~~~~~~~~~~~~~~~~~~~~~~~~~\cr
{\rm calculated}~~~~~~~~~~{\rm NA31~in~1987}~~~~~~~~~~~|~~~~~\frac{B(K_L \to \pi^0 \pi^0)}{B(K_L \to \pi^+ \pi^-)}~~~~~ \cr
|~~~~~~~~~~~~~~~~~~~~~~|~~~~~~~~~~~~~~~~~~~~~~~~|~~~~~~~~~|~~~~~~~~~~~~~~~~~~~~~~ \cr
\downarrow\,~~~~~~~~~~~~~~~~~~~\downarrow\,~~~~~~~~~~~~~~\,~~~~~~~\downarrow\,~~~~~~~|~~~~~~~~~~~~~~~~~~~~~~ \cr
\frac{B^{obs}_{\gamma\gamma}(K_L \to \mu\mu)}{B(K_L \to \pi^+ \pi^-)} =
\frac{B^{obs}_{\gamma\gamma}(K_L \to \mu\mu)}{B(K_L \to \gamma\gamma)}
\frac{B(K_L \to \gamma\gamma)}{B(K_L \to \pi^0 \pi^0)} 
\overbrace{\frac{B(K_S \to \pi^0 \pi^0 )}{B(K_S \to \pi^+ \pi^-)}
(1 - 6 Re \frac{\epsilon'}{\epsilon})} \cr
|~~~~~~~~~~~~~~~~~~~~~~~|~~~~~~~~~~~~~~~~~~~~~~~~|~~~~~~~~~~~~~~~~~~~~~|~~~~~~~~~~ \cr
1.195 \cdot 10^{-5}~~~~~~~~~~~~~~|~~~~~~~~~~~~~~~~~~~~~~~~|~~~~~~~~~~~~~~~~~~~~~|~~~~~~~~~~ \cr
0.632 \pm 0.009~~~~~~~~~~~~~~|~~~~~~~~~~~~~~~~~~~~~|~~~~~~~~~~ \cr
(2.221 \pm 0.014)^{-1}~~~~~~~~|~~~~~~~~~~ \cr
1-6(16.7 \pm 2.6)\cdot 10^{-4} \cr
\label{foo}
\eea

This yields $(3.400 \pm 0.053) \times 10^{-6}$.  A similar calculation
can be done using the results of recent determinations of the
ratio $\Gamma(2 \gamma) /\Gamma(3 \pi^0)$ from KLOE~\cite{Adinolfi:2003ca} 
and NA48~\cite{Lai:2002sr}.  For this, one also needs a value for the ratio
$B(K_L \to 3 \pi^0)/B(K_L \to \pi^+ \pi^-)$ and I use the recent KTeV results 
on $K_L$ branching ratios~\cite{Alexopoulos:2004sx}, since the correlations
between the various measurements are given. 
This yields $(3.299 \pm 0.048) \times 10^{-6}$. Averaging the $2 \pi^0$
and $3 \pi^0$ determinations yields $(3.345 \pm 0.036) \times 10^{-6}$.
The subtraction then yields $\frac{B^{disp}(K_L \to \mu^+ \mu^-)}{\bkppr} = 
(0.135 \pm 0.061) \times 10^{-6}$ (where $B^{disp}$ refers to the 
dispersive part of \bkmm).  $\frac{B^{disp}(K_L \to \mu^+ \mu^-)}{\bkppr}$
can then be multiplied by $\bkppr = (1.975 \pm 0.015) \times 10^{-3}$
~\cite{PDBook05} to obtain
$B^{disp}(K_L \to \mu^+\mu^-) = (0.267\pm 0.121) \times 10^{-9}$, or 
$B^{disp}(K_L \to \mu^+\mu^-) < 0.42 \times 10^{-9}$ at 90\% CL.  
\footnote{Note that since
\bkmm~and $B^{abs}_{\gamma\gamma}(K_L \to \mu^+ \mu^-)$ are so close,
small shifts in the component values could have relatively large consequences
for  $B^{disp}(K_L \to \mu^+\mu^-)$.}

Now if one inserts
the result of even very conservative recent CKM fits into the formula for 
the short distance part of \bkmm, one gets poor agreement with
the limit of $B^{disp}(K_L \to \mu^+\mu^-)$ derived above.  For example the
2$\sigma$ fit of the CKM Fitter Group\cite{Charles:2004jd},
$\bar\rho = 0.096 - 0.285$,
gives $B^{SD}(K_L \to \mu^+\mu^-) = (0.64-0.93)\times 10^{-9}$.  So why 
haven't we been hearing about this apparent violation of the SM?  There
are certainly viable candidates for BSM contributions to this 
decay~\cite{Isidori:2002qe,D'Ambrosio:2002fa}.

\begin{figure}[t]
\centering
\includegraphics[angle=0, height=.5\textheight]{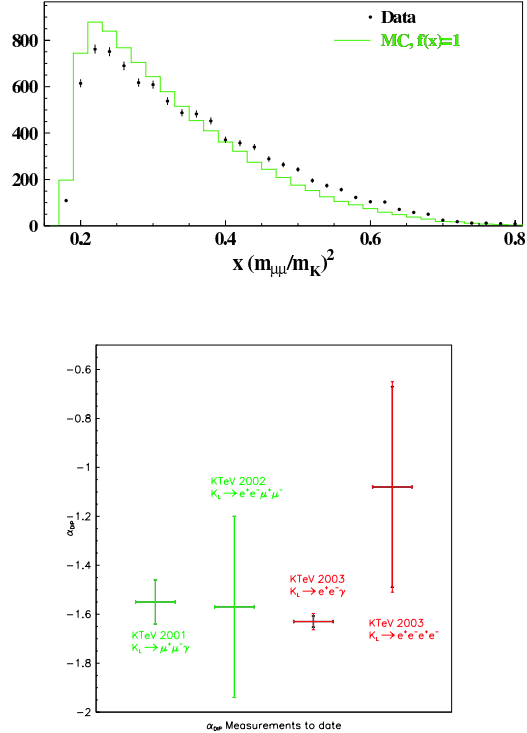}
  \caption{{\bf Top:} spectrum of $x = (m_{\mu\mu}/m_K)^2$ in 
$K_L \to \mu^+ \mu^- \gamma$ from Ref.~\protect\cite{Alavi-Harati:2001wd}.  
{\bf Bottom:} determinations of the parameter
$\alpha_{DIP}$ from four $K_L$ decays involving virtual photons.
    \label{fig:mmg} }
\end{figure}

The answer is that unfortunately $K_L \to \gamma^* \gamma^* $ also
generates a dispersive contribution, that is rather less tractable
than the absorptive part and that interferes with the
short-distance weak contribution one wants to extract.  The
problem in calculating this contribution is the necessity of including
intermediate states with virtual photons of all effective masses.
Such calculations can only be partially validated in
kaon decays containing virtual photons in the final state.  The
degree to which this validation is possible is controversial with
both optimistic~\cite{D'Ambrosio:1998jp,GomezDumm:1998gw,Isidori:2003ts}
and pessimistic~\cite{Valencia:1998xe,Knecht:1999gb} conclusions
available.  Recently
there have been talks and publications on $K_L \to \gamma e^+ e^-$
\cite{LaDue:2003xd} (93,400 events), $K_L \to \gamma \mu^+
\mu^-$\cite{Alavi-Harati:2001wd} (9327 events), $K_L \to e^+ e^- e^+
e^-$\cite{LaDue:2003xd} (1056 events), and $K_L \to \mu^+ \mu^-
e^+ e^-$\cite{Alavi-Harati:2002eh} (133 events).
Figure~\ref{fig:mmg}-top
shows the spectrum of $x = (m_{\mu\mu}/m_K)^2$ from
Ref.\cite{Alavi-Harati:2001wd}.  The disagreement between the data
(filled circles with error bars) and the prediction of pointlike
behavior (histogram) clearly indicates the presence of a form factor.  A
candidate for this is provided by the DIP
model\cite{D'Ambrosio:1998jp} which depends on parameters
$\alpha$ and $\beta$.  The latter parameter applies only in decays in which
both photons are virtual.  

Fig.~\ref{fig:mmg}-bottom shows four determinations of this parameter
from KTeV.  These are mutually consistent, but there's a disagreement
with a previous NA48 result for $K_L \to e^+ e^- \gamma$~\cite{Fanti:1999rz}.
Both the DIP and the older BMS \cite{Bergstrom:1983rj} parameterization
predict a connection between the shape
of the $\ell^+ \ell^-$ spectra and the branching ratios, that can
be exploited in the cases involving muon pairs.  In those cases 
the parameters have been determined from both the spectra and the
branching ratios.  These determinations agree at the 1-2$\,\sigma$ level.
However the available data is not yet sufficient to clearly favor either
parameterization.  In addition, it seems clear that a very large increase
in the data of processes where both photons are virtual, such as 
$K_L \to e^+ e^- \mu^+ \mu^-$ would be needed to completely test the 
DIP parameterization~\cite{Hamm:2002vy,Halkiadakis:2001fb}.
Additional effort, both experimental and theoretical, is required
before the quite precise data on \bkmm~ can be fully exploited.

	Finally, one might ask if it is possible to extract short distance
information from the decay $K_L \to e^+ e^-$.  AGS E871 observed four events
of this mode, establishing $B(K_L \to e^+ e^- ) = (8.7 {+5.7 \atop{-4.1}})
\times 10^{-12}$~\cite{Ambrose:1998cc}, the smallest 
branching ratio yet measured for an elementary particle.  
Unfortunately the SM short
distance contribution is helicity suppressed with respect to
$K_L \to \mu^+ \mu^-$ by the ratio
$m_e^2/m_{\mu}^2$ while long distance contributions
are relatively enhanced, making the extraction of SM
short distance information almost impossible
~\cite{Valencia:1998xe}.  The theoretical prediction of the
branching ratio agrees well with what is observed, which limits the
presence of BSM pseudoscalar couplings in this decay.

\subsection{\kpll}

	Like \kpnn0, the \kpll~are GIM-suppressed neutral current reactions
sensitive to short-distance SM and BSM effects, but with far different
experimental considerations.
Like \kpnn0, in the SM they are sensitive to
$Im \lambda_t$, but in general they have different sensitivity to BSM
effects~\cite{Buras:1999da}.  Although their signatures are
intrinsically superior to that of \kpnn0, they are subject to a
serious background that has no analogue in the neutral lepton case: $K_L \to
\gamma\gamma\ell^+\ell^-$.  This process, a radiative correction to
$K_L \to \gamma\ell^+\ell^-$, occurs $10^3 - 10^4$ times
more frequently than \kpll.  Kinematic cuts are
quite effective, but it is very difficult to improve the signal:background
beyond about $1:1.5$\cite{Greenlee:1990qy} and still maintain
adequate acceptance.  Both varieties of  $K_L \to
\gamma\gamma\ell^+\ell^-$ have been observed, $B(K_L \to \gamma\gamma e^+ 
e^-)_{k_{\gamma}>5 MeV} = (5.84 \pm 0.15_{stat} \pm 0.32_{syst}) \times
10^{-7}$\cite{Alavi-Harati:2000tv} and $B(K_L \to \gamma\gamma \mu^+
\mu^-)_{m_{\gamma\gamma}> 1 MeV/c^2} = (10.4 {+7.5 \atop{-5.9}}_{stat} \pm 
0.7_{syst}) \times 10^{-9}$\cite{Alavi-Harati:2000hr}; both agree
reasonably well with theoretical prediction.   By comparison, the direct 
CP-violating part of $B(K_L \to \pi^0 e^+ e^-)$, is predicted to be 
$(4.4 \pm 0.9) \times 10^{-12}$ and that of $B(K_L \to \pi^0 \mu^+ \mu^-)$ 
to be $(1.8 \pm 0.4) \times 10^{-12}$~\cite{Isidori:2004rb}.

	In addition to this background, there are two other contributions
that complicate the extraction of short-distance information from 
\kpll .  First, there is an indirect CP-violating
amplitude from the $K_1$ component of $K_L$ that is proportional to 
$\epsilon A(K_S \to \pi^0 \ell^+ \ell^-)$. 
It is of the same order of magnitude as the direct CP-violating 
amplitude and can interfere with it.  In the case of
\klpee, it yields~\cite{Isidori:2004rb}:
\be
B(K_L \to \pi^0 ee)_{CPV} \approx \left [15.7 a_S^2 \pm 6.2 a_S \frac{Im \lambda_t}{10^{-4}} + 2.4 \left(\frac{Im \lambda_t}{10^{-4}}\right)^2\right] \times 10^{-12}
\label{cpks}
\ee
where 
\be
B(K_S \to \pi^0 ee) \approx 5.2 a_S^2 \times 10^{-9}
\label{ksbr}d
\ee
For \kpmm, the corresponding expressions are~\cite{Isidori:2004rb,D'Ambrosio:1998yj}:
\be
B(K_L \to \pi^0 \mu \mu)_{CPV} \approx \left [3.7 a_S^2 \pm 1.6 a_S \frac{Im \lambda_t}{10^{-4}} + 1.0 \left(\frac{Im \lambda_t}{10^{-4}}\right)^2\right] \times 10^{-12}
\label{cpksmm}
\ee
and
\be
B(K_S \to \pi^0 \mu \mu) \approx 1.2 a_S^2 \times 10^{-9}
\label{ksbrmm}
\ee

\begin{figure}[ht]
\centering
 \includegraphics[angle=0, height=.30\textheight]{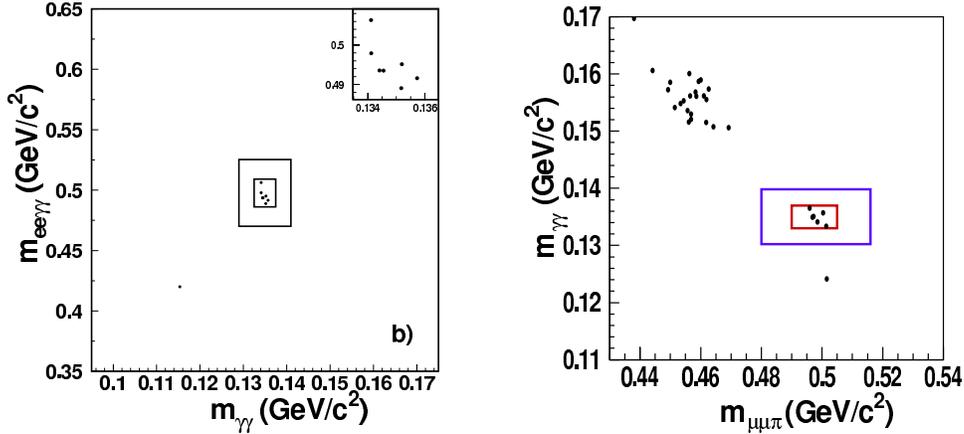}
  \caption{{\bf Left:} NA48 $K_S \to \pi^0 e^+
e^-$ candidates (from Ref~\cite{Batley:2003mu}).
{\bf Right:} NA48 $K_S \to \pi^0 \mu^+ \mu^-$ candidates (from
Ref~\cite{Batley:2004wg}).
    \label{fig:kspll} }
\end{figure}

There are now measurements for both these processes.
$B(K_S \to \pi^0 e^+ e^-)|_{m_{ee}>165} = 
(3.0 {+1.5 \atop{-1.2}} \pm 0.2) \times 10^{-9}$ from
NA48~\cite{Batley:2003mu}, based on 7 observed events with an estimated
background of 0.15 events (see Fig.~\ref{fig:kspll}).  The branching
ratio then needs to be corrected for a cut of $m_{ee}>165\,$MeV/c$^2$ that was
imposed to eliminate $K_S \to \pi^0 \pi^0$; $\pi^0 \to e^+ e^- \gamma$ events.
This yields: $B(K_S \to \pi^0 e^+ e^-) = 
(5.8 {+2.8 \atop{-2.3}} \pm 0.8) \times 10^{-9}$.
Eq.~\ref{ksbr} then yields $|a_S| = 1.06 {+0.26 \atop{-0.21}} \pm 0.07$.

NA48 has also observed six $K_S \to \pi^0 \mu^+ \mu^-$
candidates over a background of 0.22 events~\cite{Batley:2004wg} 
(see Fig.~\ref{fig:kspll}).
This yields $B(K_S \to \pi^0 \mu^+ \mu^-) = (2.9 {+1.5 \atop{-1.2}} \pm 
0.2) \times 10^{-9}$.  Eq.~\ref{ksbrmm} then gives $|a_S| = 1.54 {+0.40
\atop{-0.32}} \pm 0.06$.  Ref.~\cite{Isidori:2004rb} averages the electron
and muon results to get a best estimate of $|a_S| = 1.2 \pm 0.2$.
The ratio $\Gamma(K_S \to \pi^0 \mu^+ \mu^-)/\Gamma(K_S \to \pi^0 e^+ e^-) 
= 0.49 {+0.35 \atop{-0.29}} \pm 0.07$ is consistent
within 1 $\sigma$ with the value $\sim 0.2$ predicted in the SM (and almost
any other model).

\begin{figure}[ht]
\centering
 \includegraphics[angle=0, height=.28\textheight]{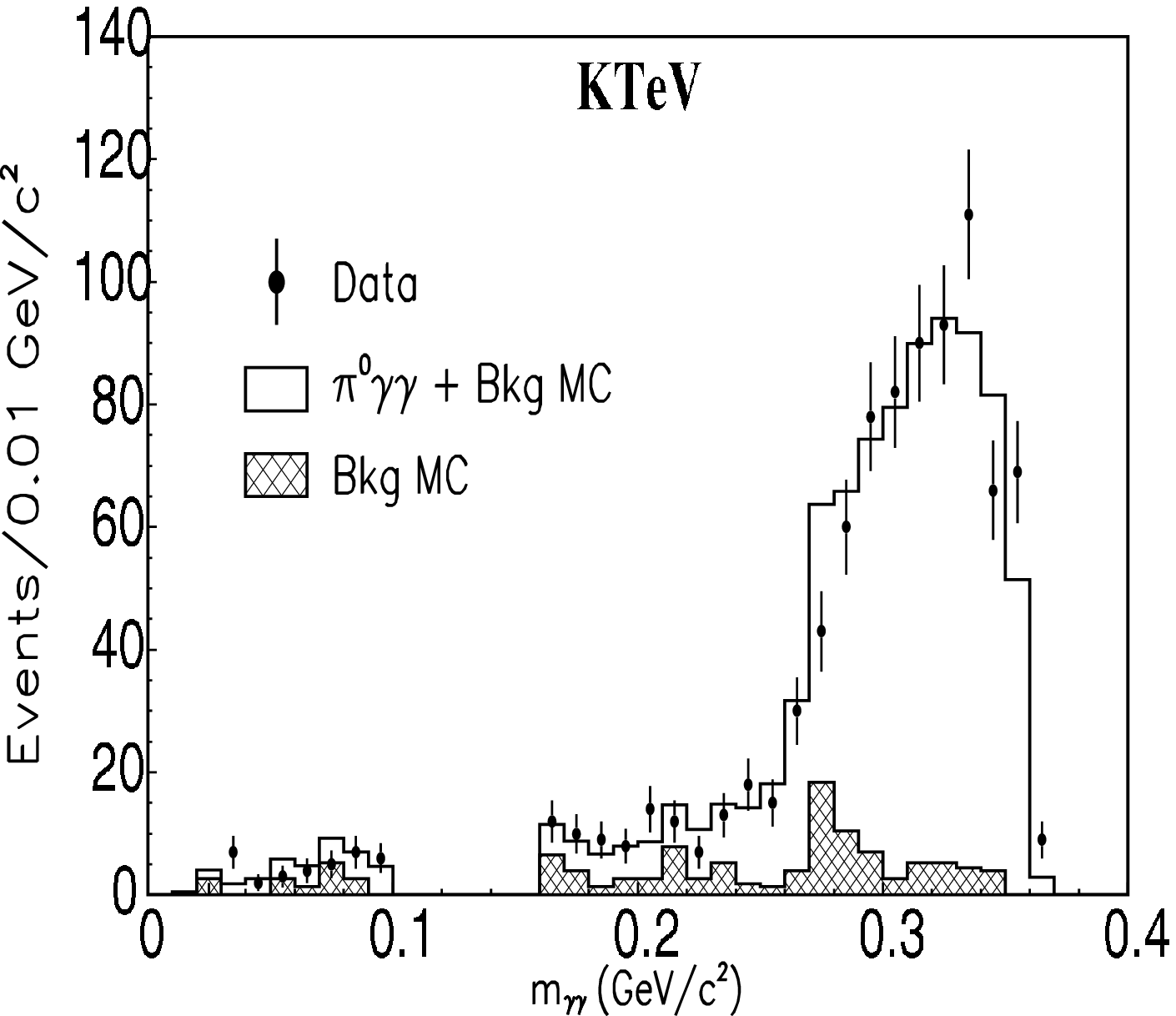}
 \includegraphics[angle=0, height=.27\textheight]{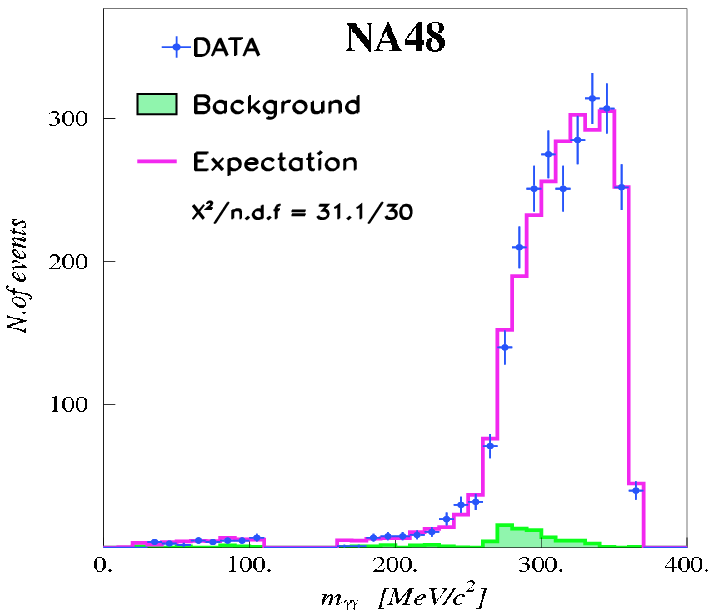}
  \caption{$m_{\gamma\gamma}$ spectrum of $K_L \to \pi^0 \gamma\gamma$
candidates from (left) KTeV~\protect\cite{Alavi-Harati:1999mu} and (right) 
NA48~\cite{Lai:2002kf}.  The striking threshold at $m_{\gamma\gamma}
\approx 2 m_{\pi}$ is due to the amplitude $A(y,z)$ in $\chi$PT (or in
pion loop models~\protect\cite{Sehgal:1972hs}).
    \label{fig:pgg0} }
\end{figure}

Another contribution of similar order, but {\it CP-conserving},
is mediated by $K_L \to \pi^0 \gamma\gamma$.  In principle
this contribution can be predicted from measurements of 
$K_L \to \pi^0 \gamma\gamma$, of which
thousands of events have been observed.  
The matrix element for this decay is given by~\cite{Ecker:1988hd}:
\bea
{\mathcal M}(K_L \to \pi^0 \gamma \gamma) = \frac{G_8 \alpha}{4 \pi}
\epsilon_{\mu}(k_1) \epsilon_{\nu}(k_2)  \big [ A(k^{\mu}_2 k^{\nu}_1 - 
k_1 \cdot k_2 g^{\mu \nu}) + \cr
 B \frac{2}{m^2_K}(p_k \cdot k_1 k^{\mu}_2 p_K^{\nu}
+p_K \cdot k_2 k^{\nu}_1 p_K^{\mu} -k_1 \cdot k_2 p_K^{\mu} p^{\nu}_K - g^{\mu\nu}
p_K \cdot k_1 p_K \cdot k_2) \big ]
\label{pggm}
\eea
where $k_1$ and $k_2$ refer to the photons. The $A$ amplitude corresponds
to $J_{\gamma\gamma} = 0$;
$B$ is a mixture of $J_{\gamma\gamma} = 0$  and $J_{\gamma\gamma} = 2$.
$G_8$ is the octet coupling constant in $\chi$PT.
Eq.~\ref{pggm} leads to 
\be 
\frac{\partial^2 \Gamma(K_L \to \pi^0 \gamma\gamma)}{\partial y \partial z} = \frac{m_K}{2^9 \pi^3}
\left [ z^2 |A + B|^2 + (y^2- \frac{1}{4}\lambda(1,r^2_{\pi},z))^2 |B|^2 \right ]
\label{pggm2}
\ee
where $z \equiv (k_1 + k_2)^2/m^2_K$, $y \equiv p_K \cdot (k_1 - k_2)/m^2_K$,
$r_{\pi} \equiv m_{\pi}/m_K$ and $\lambda(a,b,c) \equiv a^2 + b^2 + c^2
-2(ab+ac+bc)$.  Since in $K_L \to \pi^0 e^+ e^-$ the effect of $A$
is greatly suppressed by helicity conservation
and $B = 0$ at leading order in $\chi$PT, it was initially thought
that the CP-conserving contribution to $B(K_L \to \pi^0 e^+ e^-)$ would
be very small.  However the possibility of a substantial vector meson
dominance (VDM) contribution to $B$ was pointed out by 
Sehgal~\cite{Sehgal:1988ej}. Such a contribution can arise at 
$\cal{O}$$(p^6)$ in $\chi$PT.  Indeed, early measurements
of $B(K_L \to \pi^0 \gamma\gamma)$~\cite{Barr:1990hc,Papadimitriou:1991iw} 
showed that although the simple $\cal{O}$$(p^4)$ calculation was in reasonable
agreement with the $m_{\gamma\gamma}$ spectrum, it underestimated the decay 
rate 
by a factor $> 2$.  Subsequent theoretical work 
has remedied this to a large extent~\cite{Cohen:1993ta,Cappiello:1993kk,
Heiliger:1993uh,Kambor:1994tv,D'Ambrosio:1997sw}.  
Although a full $\mathcal{O}(p^6)$
calculation is not possible at present, in this work the $\mathcal{O}(p^4)$ 
calculation was  improved by ``unitarity corrections'' and the addition of a 
VDM contribution characterized by a single parameter $a_V$.  This produced
satisfactory agreement with the observed branching ratio (at least until
more precise measurements became available).  A similar 
approach~\cite{D'Ambrosio:1996zx} was successful in predicting the 
characteristics of the closely related decay $K^+ \to \pi^+ \gamma\gamma$ 
that was measured by AGS E787~\cite{Kitching:1997zj} and 
E949~\cite{Artamonov:2005ru}.  The most recent data on $K_L \to \pi^0 \gamma\gamma$
is summarized in Table~\ref{kpgg}. 
Unfortunately the two results, from
KTeV~\cite{Alavi-Harati:1999mu} and from NA48~\cite{Lai:2002kf} disagree by 
nearly $3 \sigma$ in branching ratio.
Their $m_{\gamma\gamma}$ spectra also differ rather
significantly as shown in Fig.~\ref{fig:pgg0}, leading to disagreement in 
their predictions for $B^{CP-cons}(K_L \to \pi^0 e^+ e^-)$.
This is also reflected in differing extracted values of $a_V$
seen in Table~\ref{kpees}.  As discussed below, this disagreement 
complicates assessing the prospects of extracting
$B^{direct}(K_L \to \pi^0 e^+ e^-)$ from data.

\begin{table}[htbp]
\caption{\it Recent results on \kpgg.}
\begin{tabular}{lll} \hline
Exp/Ref & $B(K_L \to \pi^0 \gamma \gamma ) \cdot 10^6$ & $a_V$ \\
\hline
KTeV~\cite{Alavi-Harati:1999mu}  & $1.68 \pm 0.07_{stat} \pm 0.08_{syst} $                & $-0.72 \pm 0.05 \pm 0.06 $              \\
NA48~\cite{Lai:2001jf}           & $1.36 \pm 0.03_{stat} \pm 0.03_{syst} \pm 0.03_{norm}$ & $-0.46 \pm 0.03 \pm 0.03 \pm 0.02_{theor} $\\
\end{tabular}
\label{kpgg}
\end{table}

In addition, the simple VDM-based formalism for predicting $B^{CP-cons}(K_L \to
\pi^0 e^+ e^-)$ from \kpgg~ was criticized
by Gabbiani and Valencia~\cite{Gabbiani:2001zn}. They pointed out
that the use of a single parameter $a_V$ can artificially correlate the 
$A$ and $B$ amplitudes.  They showed that a three parameter
expression inspired by $\mathcal{O}(p^6)$ $\chi$PT fits the KTeV data
quite as well as the conventional one based on $a_V$ and gives a
significantly different prediction for $B^{CP-cons}(K_L \to
\pi^0 e^+ e^-)$ as seen in Table~\ref{kpees}.  In a subsequent paper
~\cite{Gabbiani:2002bk}, they find they can make a good
simultaneous fit to the NA48 decay spectrum and rate using the same
technique and show that this is not possible using just $a_V$.  Not
surprisingly, their
predictions of their three parameter fits for $ B^{CP-cons}(K_L \to
\pi^0 e^+ e^-)$ also differ markedly between the two experiments as
shown in Table~\ref{kpees}.

	Finally there are significant uncertainties in the extraction
of the dispersive contribution to
$ B^{CP-cons}(K_L \to \pi^0 e^+ e^-)$~\cite{Donoghue:1995yt, Gabbiani:2001},
which is similar in size to the absorptive contribution and so not at
all negligible.

	Subsequently Buchalla, D'Ambrosio and
Isidori\cite{Buchalla:2003sj} took a different approach to analyzing
the NA48 data.  They used the same three parameters but applied a
constraint given by the NA48 measurement of $B(K_S \to
\gamma\gamma)$\cite{Lai:2002sr}.  They obtained somewhat different
values for the parameters but agreed that the CP-conserving component
of $K_L \to \pi^0 e^+ e^-$ should be small ($< 3 \times
10^{-12}$).  In fact for similar assumptions on the ratio of dispersive
to absorptive contribution and using the data in the same way, the two approaches
yield very similar results.  For example,
using the central value of the NA48 measurement of the
low $m_{\gamma\gamma}$ part of the $K_L \to \pi^0 \gamma\gamma$ 
spectrum and the recipe for the dispersive part of the contribution of 
Ref.~\cite{Donoghue:1995yt}, the method of Buchalla {\it et al.} yields 
the value listed in Table~\ref{kpees}.  Based on the NA48
measurement, a reasonable estimate of $B^{CP-cons}(K_L \to \pi^0 e^+ e^-)$
is $(4\pm 4) \times 10^{-13}$.

\begin{table}[htbp]
\caption{\it Predictions for $B^{CP-cons}(K_L \to \pi^0 e^+ e^-)$.}
\begin{tabular}{cccc} \hline
Exp. &  $a_V$ fit from  &  3 parameter fit by &  3 parameter fit by \\
     & experimental paper  &  Gabbiani \& Valencia &  Buchalla {\it et al.}\\
\hline
KTeV  &  $(1.0-2.0) \cdot 10^{-12}$  &  $7.3 \cdot 10^{-12}$ & \\
NA48  &  $(0.47 {+0.22 \atop -0.18}) \cdot 10^{-12}$ &$(0.46 {+0.22 \atop -0.17}) \cdot 10^{-12}$ & $0.34 \times 10^{-12}$ \\
\end{tabular}
\label{kpees}
\end{table}

To summarize, there is reasonable agreement on how to extract the absorptive
contribution to $B^{CP-cons}(K_L \to \pi^0 e^+ e^-)$ from the observed low 
$m_{\gamma\gamma}$ part of the $K_L \to \pi^0 \gamma\gamma$ branching ratio.
There is some uncertainty over the dispersive part of the contribution, but
this is immaterial if the NA48 result is correct.  However, if the KTeV result
is the correct one, $B^{CP-cons}(K_L \to \pi^0 e^+ e^-)$ could be a 
non-negligible
part of the total and would complicate the extraction of short-distance
information from results on this process.  Thus further experimental work on
$K_L \to \pi^0 \gamma\gamma$ would be very welcome.

	The situation in the muonic case is quite different.  Here there is 
very little 
helicity suppression of the $J_{\gamma\gamma}=0$ part of the CP-conserving
contribution to $B(K_L \to \pi^0 \mu^+ \mu^-)$ and this is in fact expected to 
dominate the
contribution.  Recent work by Isidori {\it et al.}~\cite{Isidori:2004rb} has
shown that although there are considerable uncertainties in the absolute 
calculation of the CP-conserving contribution, the ratio 
$B^{CP-cons}(K_L \to \pi^0 \mu^+ \mu^-)/B(K_L \to \pi^0 \gamma\gamma)$  
should be calculable to $\sim$30\%.  Working from the average of the
KTeV and NA48 results for $B(K_L \to \pi^0 \gamma\gamma)$, they obtain
$B^{CP-cons}(K_L \to \pi^0 \mu^+ \mu^-) = (5.2 \pm 1.6) \times
10^{-12}$.

	Plugging $a_S=1.2 \pm 0.2$ and Im$\lambda_t = (1.36 \pm 0.12) \times
10^{-4}$ into Eqs.~\ref{cpks} and \ref{cpksmm}.
\bea
B(K_L \to \pi^0 ee)_{CPV} \approx &  [(22.6 \pm 7.5)_{mix} \pm (10.1 \pm 2.0)_{int} + \nonumber \\
&(4.44 \pm 0.87)_{dir}] \times 10^{-12}
\label{cpksval}
\eea
and
\bea
B(K_L \to \pi^0 \mu \mu)_{CPV} \approx & [(5.3 \pm 1.8)_{mix} \pm (2.6 \pm 0.5)_{int} + \nonumber \\
&(1.8 \pm 0.4)_{dir}]\times 10^{-12}
\label{cpksmmval}
\eea

To obtain a prediction for the complete branching ratios, one
needs to fix the sign of the interference between direct and indirect
CP-violation, and decide what to take for the CP-conserving part.
Positive interference has been preferred by theorists from the earliest days of
this subject\cite{Gilman:1979ud}\cite{Dib:1988js}, and up-to-date arguments
are given in Ref.~\cite{Buchalla:2003sj}.  If we accept this choice,
Eqs.~\ref{cpksval} and ~\ref{cpksmmval} give $B(K_L \to \pi^0 ee)_{CPV} 
\approx  (3.7 \pm 0.8) \times 10^{-11}$ and
$B(K_L \to \pi^0 \mu \mu)_{CPV} \approx (1.0 \pm 0.2) \times 10^{-11}$
respectively.  The CP-conserving contribution is a problem for the electron
mode.  If we use the NA48 result, from the above discussion we 
estimate  $(4 \pm 4) \times 10^{-13}$ for this.  If instead, we accept
the KTeV result, to be conservative we should use a much larger value,
$(4.0 \pm 3.5) \times 10^{-12}$.  These give:
\bea
B(K_L \to \pi^0 e^+ e^-)_{\rm NA48} \approx & (3.8 \pm 0.8)\times 10^{-11} \\
B(K_L \to \pi^0 e^+ e^-)_{\rm KTeV} \approx & (4.1 \pm 0.9)\times 10^{-11}.
\label{totale}
\eea
In the muonic mode, the CP-conserving contribution is much less problematic,
and using the value discussed above, one gets:
\be
B(K_L \to \pi^0 \mu^+ \mu^-) \approx (1.5 \pm 0.3) \times 10^{-11}
\label{totalm}
\ee

\begin{figure}[ht]
\centering
 \includegraphics[angle=0, height=.30\textheight]{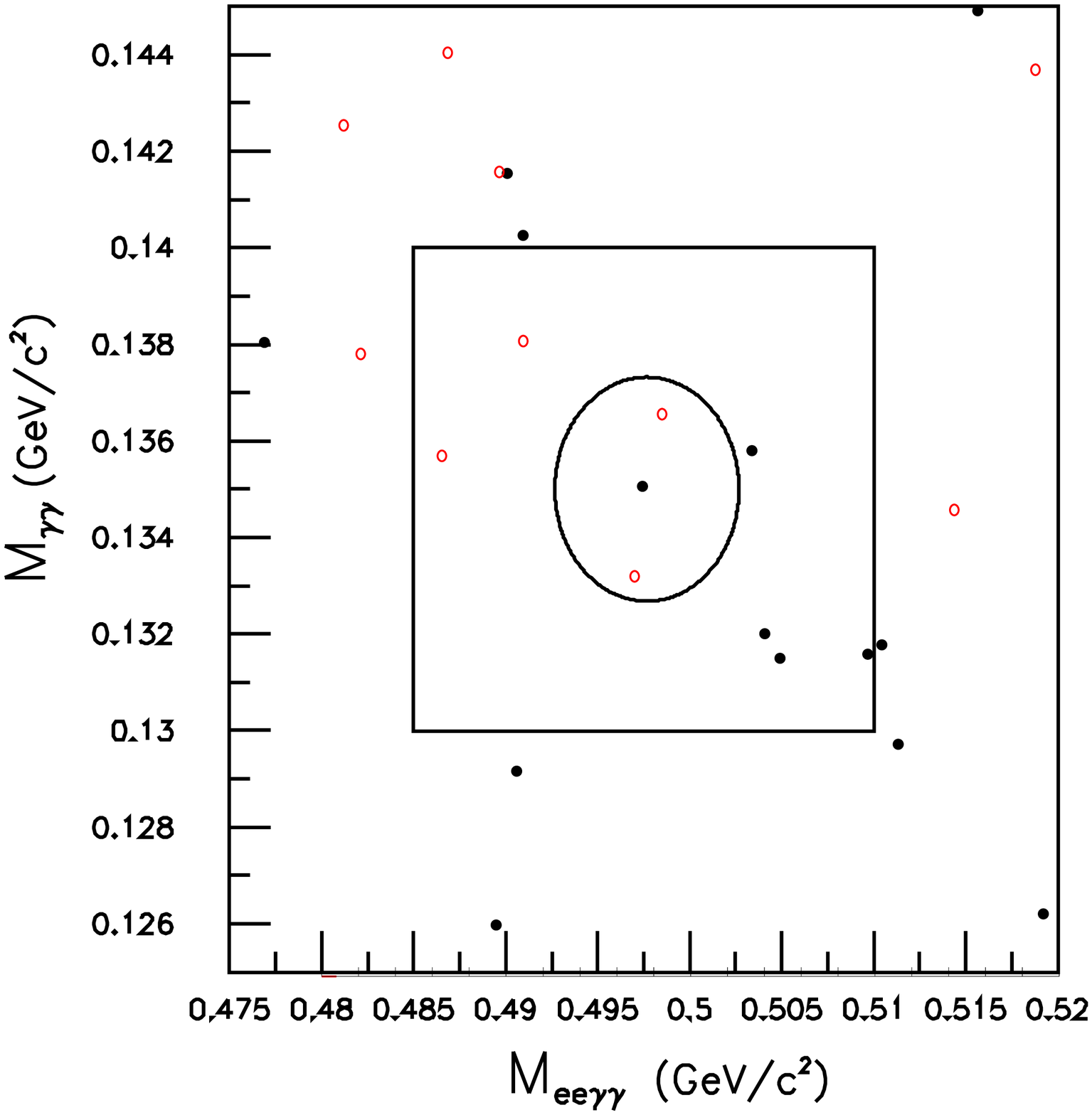}
 \includegraphics[angle=0, height=.325\textheight]{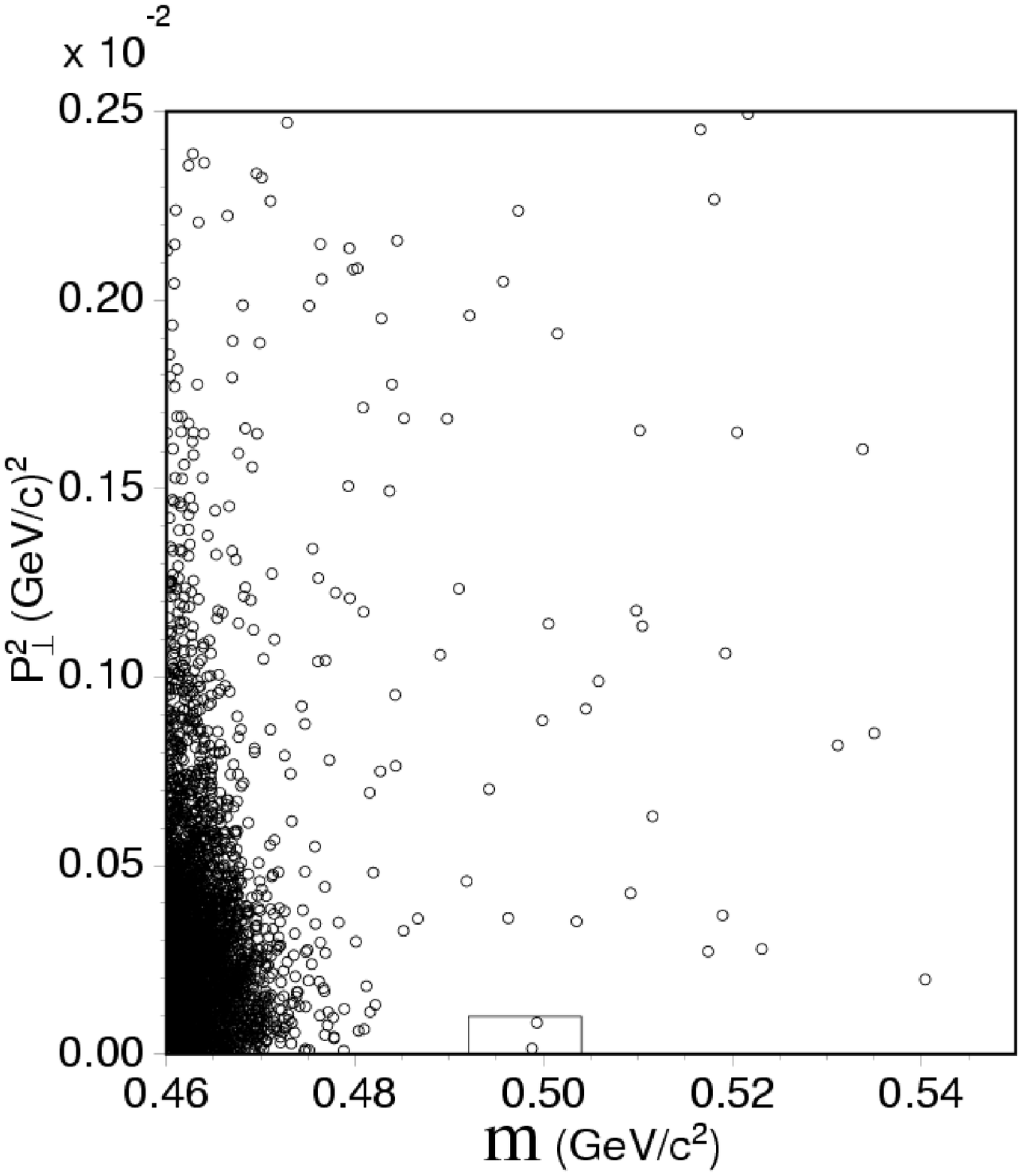}
  \caption{Signal planes showing candidates for $K_L \to \pi^0 e^+ e^-$ (left
from Ref.~\protect\cite{Alavi-Harati:2000sk} and
~\protect\cite{Alavi-Harati:2003mr}) and $K_L \to \pi^0 \mu^+ \mu^-$ (right
from Ref.~\protect\cite{Alavi-Harati:2000hs}).
    \label{fig:llpi0} }
\end{figure}

	The current experimental status of \kpll~is summarized in Table
~\ref{Kpll} and Fig.\ref{fig:llpi0}.  A factor $\sim 1.3$ more 
$K_L \to \pi^0 \mu^+ \mu^-$ data is expected from the KTeV
1999 run. 

\begin{table}[h]
\caption{\it Results on \kpll.}
\begin{tabular}{lllll} \hline
Mode & 90\% CL upper limit & Est. bkgnd.& Obs. evts. &Ref. \\
\hline
$K_L \to \pi^0 e^+ e^-$     & $2.8 \times 10^{-10}$ & $2.05\pm 0.54$ & 3 & \cite{Alavi-Harati:2003mr} \\
$K_L \to \pi^0 \mu^+ \mu^-$ & $3.8 \times 10^{-10}$ & $0.87\pm 0.15$ & 2 & \cite{Alavi-Harati:2000hs} \\
\end{tabular}
\label{Kpll}
\end{table}

As can be seen from the table and figure, background in both modes is
already starting to be observed at a sensitivity an order of magnitude
short of the expected signal level.  The problems of extracting
a value of Im$\lambda_t$ from these modes have been discussed in
Ref.~\cite{Littenberg:2002um} among other places, and summaries
of various schemes to deal with these problems are given in
Refs.~\cite{Buchalla:2003sj} and \cite{Isidori:2004rb}.  

Although the situation appears difficult, it's worthwhile taking a
closer look:  
\begin{enumerate}
\item The observation of $K_S \to \pi^0 \ell^+
\ell^-$ with higher than expected rates has made it
likely that the SM branching ratios of the $K_L$ modes are also
higher than previously believed.  
\\
\item If we compare the single event
sensitivity of the recent data with that of the
residual $K_L \to \gamma\gamma \ell^+ \ell^-$ which is the least
tractable background, the ratio is much less than an order of
magnitude.  In their 1999 $K_L \to \pi^0 ee$ run, KTeV had a single
event sensitivity of $1.04 \times 10^{-10}$ and an estimated 
residual $K_L \to \gamma\gamma ee$ background of $0.99 \pm 0.35$
events\cite{Alavi-Harati:2003mr}.  This implies a S:B (signal to
background) of 1:2.5.  In their 1997 $K_L \to \pi^0 \mu\mu$
data, KTeV had a single event sensitivity of $0.75 \times 10^{-11}$ and
a calculated residual $K_L \to \gamma\gamma \mu\mu$ background of
$0.37 \pm 0.03$ events~\cite{Alavi-Harati:2000hs}\footnote{They
observed a total of 2 (presumably background) events.  I am assuming
the backgrounds other than $K_L \to \gamma\gamma\mu\mu$ can be
suppressed in a future experiment.}.  This implies S:B = 1:1.9. 
\\
\item The interest in $K_L \to \pi^0 \ell^+ \ell^-$ has evolved from
being a possible source of information on Im$\lambda_t$, to its
role as  an arena for probing BSM effects.
\end{enumerate}

\begin{figure}[ht]
\centering
 \includegraphics[angle=0, height=.25\textheight]{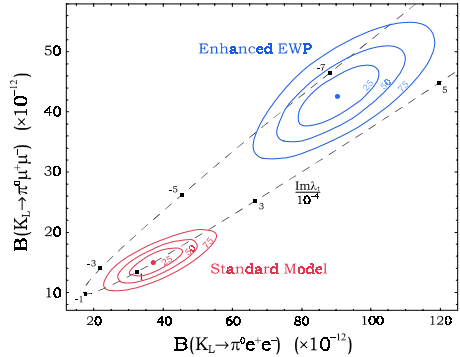}
  \caption{Relationship between $B(K_L \to \pi^0 \mu^+ \mu^-)$ and
$B(K_L \to \pi^0 e^+ e^-)$ for the SM and for a model that introduces new
physics (see text for details). The 25\%, 50\%, and 75\% CL 
regions (due to theoretical and parametric uncertainties)
are shown for each case.  From Ref.~\cite{Isidori:2004rb}.
    \label{fig:isuex} }
\end{figure}

With this in mind one can ask, for example, what single event
sensitivity would be needed to establish a factor two effect at
3$\sigma$ in these modes, given the background levels mentioned
above.  For the electronic case, the answer is
$10^{-12}$; for the muonic case, $0.4 \times 10^{-12}$.  In fact this
sensitivity for the latter mode could have been reached by the KaMI
experiment at Fermilab~\cite{Alexopoulos:2001}, had it gone forward,
in about 3 years of running.  If instead, the SM were correct, a 30\%
measurement of the BR would have resulted.  Thus a next-generation
experiment could make very useful measurements.  Another example, given
by Buchalla {\it et al.}, is illustrated in Fig.~\ref{fig:isuex}.
It shows $B(K_L \to \pi^0 \mu^+ \mu^-)$ vs $B(K_L \to \pi^0 e^+ e^-)$ 
for the SM, and that for a recent model with enhanced electroweak penguins
~\cite{Buras:2004ub} designed to explain apparent anomalies in hadronic
$B$ decays.  In this BSM scenario, a $K_L \to \pi^0 \mu^+ \mu^-$
experiment of $0.4 \times 10^{-12}$/event statistical sensitivity would observe 
$180 \pm 13.4$ events of which 110 would be signal, to be compared to 37 signal
events expected for the SM.  

\section{Conclusions}

Lepton flavor violation experiments have probed sensitivities
corresponding to mass scales of well over 100 TeV, making life
difficult for models predicting accessible LFV in kaon decay and
discouraging new dedicated experiments of this type.

        The existing precision measurement of \kmm\ will be very
useful if theorists can make enough progress on calculating the
dispersive long-distance amplitude, perhaps helped by experimental
progress in $K_L \to \gamma \ell^+ \ell^-$, $K_L \to$ 4~leptons, etc.
The exploitation of \kmm~ would also be aided by higher precision
measurements of the branching ratios of some of the normalizing
reactions, such as $K_L \to \gamma\gamma$.

        \kpnnp\ should clearly be further exploited.  Two initiatives
are devoted to reaching the $10^{-12}$/event level: the P326 in-flight
proposal to CERN, and an LOI for an advanced stopped-$K^+$ experiment
at J-PARC.  The first dedicated experiment to seek \kpnn0~(E391a) is
proceeding and there are plans to continue the pursuit of this
reaction at J-PARC with the eventual goal of making a $\leq 10\%$ measurement of
the branching ratio.

	Measurements of \kpnnp~and \kpnn0~can determine an alternative
unitarity triangle that will offer a critical comparison with results
from the $B$ system.  If new physics is in play in the flavor sector,
the two triangles will almost certainly disagree.  Moreover these
reactions can be calculated very precisely so that they can elucidate
flavor couplings in almost any BSM theory.

	There are no near-term plans to pursue $K_L \to \pi^0 \ell^+
\ell^-$ although the recent observation of $K_S \to \pi^0 \ell^+
\ell^-$ indicates that these reactions may be much more tractable than
previously believed.  They can probe BSM operators not
accessible to $K \to \pi \nu\bar\nu$, so it is unfortunate that they will
probably not be pursued unless a BSM signal is seen in the latter.

I thank D. Bryman, A. Buras, A. Ceccucci, T. Inagaki, G. Isidori,
S. Kettell, T. Komatsubara, W. Marciano, R. Shrock, G. Valencia, and
M. Zeller for useful discussions, access to results, and other
materials. This work was supported by the U.S.  Department of Energy
under Contract No. DE-AC02-98CH10886.


\bibliographystyle{laurie}
\bibliography{varenna}

\end{document}